\documentclass[a4paper,12pt]{article}
\usepackage{amsfonts}
\usepackage{amssymb}
\usepackage{amsmath}
\usepackage{url}

\numberwithin{equation}{section}
\def\be{\begin{equation}}
\def\ee{\end{equation}}
\def\ba{\begin{array}}
\def\ea{\end{array}}

\newcommand{\bea}{\begin{eqnarray}}
\newcommand{\eea}{\end{eqnarray}}

\begin{document}

\begin{titlepage}

     \thispagestyle{empty}
    \begin{flushright}
        \hfill {CERN-PH-TH/2013-072} \\
        %\hfill {IP-TH-../13} \\

    \end{flushright}

    %\vspace{2pt}

    \begin{center}

         {\huge{\textbf{Dualities Near the Horizon}}}
         \vspace{20pt}
        % \vspace{45pt}

{\Large{{\bf Sergio Ferrara$^{1,2}$}, {\bf Alessio Marrani$^3$}, \\{\bf Emanuele Orazi$^{4}$}, {\bf Mario Trigiante$^{5}$}}}

        \vspace{5pt}

        {$1$ Physics Department, Theory Unit, CERN,\\CH 1211, Geneva 23, Switzerland\\
        \texttt{sergio.ferrara@cern.ch}}

         \vspace{5pt}

         {$^2$ INFN - Laboratori Nazionali di Frascati,\\Via Enrico Fermi
40, I-00044 Frascati, Italy}

\vspace{5pt}

{$^3$ Instituut voor Theoretische Fysica, KU Leuven,\\
Celestijnenlaan 200D, B-3001 Leuven, Belgium\\
\texttt{alessio.marrani@fys.kuleuven.be}}

 \vspace{5pt}

 {$^4$ International Institute of Physics,\\Federal University of
Rio Grande do Norte,\\Av. Odilon Gomes de Lima 1722, Capim Macio,\\Natal-RN 59078-4
00, Brazil\\\texttt{orazi.emanuele@gmail.com}}

 \vspace{5pt}

        {$^5$ Dipartimento di Fisica, Politecnico di Torino,\\
        Corso Duca degli Abruzzi 24, I-10129 Torino, Italy\\
        and INFN - Sezione di Torino, Italy\\
        \texttt{mario.trigiante@polito.it}}

        \vspace{10pt}
\end{center}

\vspace{10pt}

\begin{abstract}
In $4$-dimensional supergravity theories, covariant under symplectic
electric-magnetic duality rotations, a significant role is played by
the symplectic matrix $\mathcal{M}(\varphi)$, related to the
coupling of scalars $\varphi$ to vector field-strengths. In
particular, this matrix enters the twisted self-duality condition
for $2$-form field strengths in the symplectic formulation of
generalized Maxwell equations in the presence of scalar fields.

In this investigation, we compute several properties of this matrix in
relation to the attractor mechanism of extremal (asymptotically flat) black
holes. At the attractor points with no flat directions (as in the $\mathcal{N%
}=2$ BPS case), this matrix enjoys a universal form in terms of the
dyonic charge vector $\mathcal{Q}$ and the invariants of the
corresponding symplectic representation $\mathbf{R}_{\mathcal{Q}}$
of the duality group $G$, whenever the scalar manifold is a
symmetric space with $G$ simple and \emph{non-degenerate of type ${\rm
E}_7$}.

At attractors with flat directions, $\mathcal{M}$ still depends on flat
directions, but not $\mathcal{MQ}$, defining the so-called Freudenthal dual
of $\mathcal{Q}$ itself. This allows for a universal expression of the
symplectic vector field strengths in terms of $\mathcal{Q}$, in the
near-horizon Bertotti-Robinson black hole geometry.

\end{abstract}

\end{titlepage}
\newpage \tableofcontents

%%%%%%%%%%%%%%%%%%%%%%%%%%%%%%%%%%%%%%%%%%%%%%%%%%%%%%%%%%%%%%%%%%%%%%

\section{\label{Intro}Introduction}

One of the most appealing properties of extended (ungauged) four-dimensional
supergravities (\textit{i.e.} locally supersymmetric models with no less
than $8$ supercharges) is their on-shell global symmetry which is
conjectured to encode the known string/M-theory dualities \cite{HT-1}. The
corresponding global symmetry group $G$, to be also dubbed $U$-duality, is
the isometry group of the scalar manifold (\textit{i.e.}, global symmetry of
the scalar field sigma-model), whose (non-linear) action on the scalar
fields is combined with a linear symplectic action on the $n$ electric field
strengths $F_{\mu \nu }^{\Lambda }$, $\Lambda =0,\dots ,n-1$, and their
magnetic duals $G_{\Lambda \mid \,\mu \nu }$ \cite{GZ} (electric-magnetic
duality action of $G$). The latter action being defined by an embedding of $%
G $ in the symplectic group $\mathrm{Sp}(2n,\mathbb{R})$, so that $F_{\mu
\nu }^{\Lambda }$, together with $G_{\Lambda \,\mu \nu }$, transform under
electric-magnetic duality in a symplectic representation $\mathbf{R}_{%
\mathcal{Q}}$ of $G$. This embedding, which determines the couplings of the
vector fields to all the other fields in the action, is built-in the
definition of a flat symplectic bundle over the scalar manifold, which is a
common mathematical feature of $\mathcal{N}\geqslant 2$-extended
supergravities \cite{strom,ADF-revisited,FK-N=8}.

Solutions to these theories naturally arrange themselves in orbits with
respect to the action of $G$, and important physical properties are captured
by $G$-invariant quantities. A notable example are the extremal, static,
asymptotically-flat black holes in $D=4$, which have deserved considerable
attention in the literature during the last $20$ years or so, since they
provide a valuable tool for studying string/M-theory dualities. These
solutions are naturally coupled to scalar fields as a consequence of the
non-minimal couplings of these to the vector fields in the supergravity
action. Near the horizon, however, they exhibit an \emph{attractor mechanism}
\cite{AM,FGK}: the near-horizon geometry, which is described by an $%
AdS_{2}\times S^{2}$ Bertotti-Robinson space-time \cite{BR}, is independent
of the values of the scalar fields at radial infinity, and only depends on
the quantized magnetic and electric charges $p^{\Lambda },\,q_{\Lambda }$.
In particular the horizon area $A_{H}$, which is related to the entropy $S$
of the solution through the Bekenstein- Hawking formula \cite{Bek-Haw}, is
expressed in terms of the \emph{quartic} invariant $I_{4}(p,q)$ of the
representation $\mathbf{R}_{\mathcal{Q}}$ of $G$, only depending on $%
p^{\Lambda },\,q_{\Lambda }$ (we set $8\pi G_{N}=c=\hbar =1$):
\begin{equation}
S=\frac{A_{H}}{4}=\pi \,\sqrt{|I_{4}(p,q)|}\,.
\end{equation}%
This is a consequence of the fact that the horizon represents an
asymptotically stable equilibrium point for the radial evolution of those
scalar fields which are effectively coupled to the solution and thus affect
its geometry. In other words, such scalars flow from radial infinity to the
horizon toward values which only depend on the quantized charges (fixed
values). The horizon fixed point is defined by extremizing an effective
potential $V_{BH}(\varphi ;p,q)$ ($\varphi $ generically denoting the scalar
fields) \cite{FGK}:
\begin{equation}
V_{BH}\left( \varphi ,\mathcal{Q}\right) :=-\frac{1}{2}\mathcal{Q}^{T}%
\mathcal{M}\left( \varphi \right) \mathcal{Q},  \label{V_BH}
\end{equation}%
where $\mathcal{Q}=(p^{\Lambda },\,q_{\Lambda })$ is the vector quantized
charges in the representation $\mathbf{R}_{\mathcal{Q}}$ of $G$. The value
of this potential at the horizon defines its area, being equal to $\sqrt{%
|I_{4}(p,q)|}$. The scalar fields which are not fixed at the horizon are
those which are not effectively coupled to the black hole charges, and they
are flat directions of $V_{BH}$. They will be denoted by $\varphi _{flat}$.
In the above formula, $\mathcal{M}(\varphi )$ is a $2n\times 2n$ symmetric,
symplectic, negative-definite matrix-valued function of the scalar fields.
In all extended supergravities it is defined by the flat symplectic bundle
over the scalar manifold. In fact, it encodes all the information about the
non-minimal couplings of the scalar to the vector fields in the action
through the kinetic term of the latter and the generalized theta-term.
Moreover it allows to define the so called \emph{Freudenthal duality} \cite%
{Duff-FD}, extensively studied in \cite{FMY-FD-1,Ort-1,F-dual-L}, which we
shall be dealing with in the following.

An interesting question to be posed is what happens to the geometric
structures associated with the scalar manifold, \textit{e.g.} pertaining to
its symplectic bundle, near the horizon. In the present investigation, we
focus on the matrix $\mathcal{M}(\varphi )$, because of its relevance to the
geometry of the supergravity model.

At the horizon $\mathcal{M}(\varphi )$ depends on $\mathcal{Q}$, through the
fixed scalars, and on the flat directions:
\begin{equation}
\left. \mathcal{M}(\varphi )\right\vert _{\mathrm{horizon}}=\mathcal{M}^{H}(%
\mathcal{Q},\,\varphi _{flat})\,.
\end{equation}%
The dependence on the flat directions drops out already when we contract $%
\mathcal{M}^{H}$ once with the charge vector. This implies the independence
of the vector field-strengths at the horizon from $\varphi _{flat}$.

On general grounds, using the properties of $\mathcal{M}(\varphi )$, one can
show that if we act on the solution by means of an element $g$ of $G$, which
maps $\varphi $ into $\varphi ^{\prime }$ and $\mathcal{Q}$ into $\mathcal{Q}%
^{\prime }$, the matrix $\mathcal{M}(\varphi )$ at the horizon transforms as
follows:\footnote{%
Here and in the following we use the short-hand notation $g^{-T}:=(g^{-1})^T$%
.}
\begin{equation}
\mathcal{M}^{H}(\mathcal{Q}^{\prime },\,\varphi _{flat}^{\prime }\,)=g^{-T}%
\mathcal{M}^{H}(\mathcal{Q},\,\varphi _{flat})\,g^{-1}\,,  \label{rell-1}
\end{equation}%
where, with an abuse of notation, we have denoted by $g$ also the symplectic
$2n\times 2n$ matrix representing the corresponding $G$-element on
contravariant vectors of $\mathbf{R}_{\mathcal{Q}}$.

In the absence of flat directions, the above equation suggests that $%
\mathcal{M}^{H}(\mathcal{Q})$ should be described in terms a symmetric,
symplectic, negative-definite matrix defined on the $G$-orbit of $\mathcal{Q}
$, and thus constructed out of $\mathcal{Q}$ and of $G$--invariant tensors.
Restricting our analysis to symmetric models with group $G$ simple of ``type
$\mathrm{E}_7$'' \cite{brown}(with the exclusion of the \emph{degenerate
cases}, see footnote 7 below), for charge vectors $\mathcal{Q}$ with $I_{4}(%
\mathcal{Q})>0$ we could construct such a matrix $M(\mathcal{Q})$ using a
simple \textit{Ansatz}, which involves only $\mathcal{Q}$ and $G$-invariant
tensors, and imposing the following properties of $\mathcal{M}^{H}$:
\begin{equation}
\begin{array}{l}
M\mathbb{C}M=\mathbb{C}\,\,;\, \\
\\
M\mathcal{Q}=-\frac{\epsilon }{2\sqrt{\left\vert I_{4}\right\vert }}\frac{%
\partial I_{4}}{\partial \mathcal{Q}},%
\end{array}
\label{masterequation!}
\end{equation}%
where $I_{4}=:\epsilon \left\vert I_{4}\right\vert $, and $\mathbb{C}$ is
the symplectic invariant $2n\times 2n$ antisymmetric matrix. \footnote{%
Note that the second of (\ref{masterequation!}) \cite{FMY-FD-1} implies%
\begin{equation}
\mathcal{Q}^{T}\,M\,\mathcal{Q}=-2\,\sqrt{|I_{4}(\mathcal{Q})|}\,;
\label{masterequation!2}
\end{equation}%
however, it can be checked that this yields the same condition (namely, (\ref%
{I4cond}) further below) on the real coefficients $A$, $B$ and $C$ of the
\textit{Ansatz} (\ref{Ans-3})-(\ref{Ans-4}).} Starting from the same general
\textit{Ansatz} we actually find two solutions to the above equations,
denoted by $M_{+}(\mathcal{Q})$ and $M_{-}(\mathcal{Q})$. For charges with $%
I_{4}(\mathcal{Q})>0$ and no flat directions, we give arguments in favor of
the identification of one of these matrices ($M_{+}$) with $\mathcal{M}^{H}(%
\mathcal{Q})$. The other solution ($M_{-}$), on the other hand, is never
negative definite and has the general form:
\begin{equation}
M_{-,MN}=-\frac{\partial ^{2}\sqrt{|I_{4}(\mathcal{Q})|}}{\partial \mathcal{Q%
}^{M}\partial \mathcal{Q}^{N}}\,.  \label{introHess}
\end{equation}%
This Hessian has been considered in the literature, see \cite%
{Ferrara:2006js,LG-2}, though in different contexts.

As far as regular BPS solutions in $\mathcal{N}=2$ supergravities are
concerned, the two matrices $M_{\pm }$ enjoy an interesting interpretation
as the value at the horizon of two characteristic symplectic, symmetric
matrices of the theories: the matrix $\mathcal{M}$ which is constructed out
of the real and imaginary parts of the period matrix $\mathcal{N}_{\Lambda
\Sigma }(\varphi )$ (defining the generalized theta-term and the kinetic
term for the vector fields, respectively), and a matrix $\mathcal{M}^{(F)}$,
constructed just as $\mathcal{M}$, but in terms of the real and imaginary
parts of a different complex matrix, namely the Hessian $\mathcal{F}%
_{\Lambda \Sigma }$ of the holomorphic prepotential of the special K\"{a}%
hler manifold. In terms of the covariantly holomorphic section $V=(V^{M})$
of the special K\"{a}hler manifold describing the vector multiplet scalars $%
z^{i}$, and of its covariant derivatives $U_{i}=D_{i}V=(U_{i}^{M})$ (we use
the notations of \cite{Andrianopoli:1996cm}), the two matrices have the
following expressions:
\begin{align}
\mathcal{M}(z,\bar{z})& =\mathbb{C}\left( V\bar{V}^{T}+\bar{V}%
V^{T}+U_{i}\,g^{i\bar{\jmath}}\bar{U}_{\bar{\jmath}}^{T}+\bar{U}_{\bar{\jmath%
}}g^{\bar{\jmath}i}U_{i}^{T}\right) \mathbb{C}\,, \\
\mathcal{M}^{(F)}(z,\bar{z})& =\mathbb{C}\left( V\bar{V}^{T}+\bar{V}%
V^{T}-U_{i}\,g^{i\bar{\jmath}}\bar{U}_{\bar{\jmath}}^{T}-\bar{U}_{\bar{\jmath%
}}g^{\bar{\jmath}i}U_{i}^{T}\right) \mathbb{C}\,.
\end{align}%
The former was given in \cite{FK-N=8} and \cite{Andrianopoli:2009je}, and it
is the real part of the identity (1.16) of \cite{FMY-FD-1}. On the other
hand, the latter expression follows from (1.13) of \cite{Ort-1};
furthermore, $\mathcal{Q}^{T}\mathcal{M}^{(F)}(z,\bar{z})\mathcal{Q}$ agrees
with Eq. (57) of \cite{Ceresole:1995ca}. In $\mathcal{N}\geqslant 2$%
-extended supergravities, for charge orbits characterized by $I_{4}(\mathcal{%
Q})<0$, the two matrices $M_{\pm }$, though still satisfying the second of (%
\ref{masterequation!}), are \textit{anti-symplectic}, namely for them the
following property holds:%
\begin{equation}
M_{\pm }\mathbb{C}M_{\pm }=-\mathbb{C}.  \label{anti-sympl}
\end{equation}%
The matrix $M_{+}$, in particular, for all regular charge-orbits, as opposed
to $M_{-}$, has the notable property of being an \emph{automorphism} of the
U-duality algebra $\mathfrak{g}$, that is $\mathfrak{g}$, in the
representation $\mathbf{R}_{\mathcal{Q}}$, is invariant under the adjoint
action of $M_{+}$ (if $I_{4}<0$, being $M_{+}$ anti-symplectic, will be
characterized as an \emph{outer} automorphism). On the other hand $M_-$ is
still, in all regular orbits, identified with the Hessian (\ref{introHess}).
Moreover both $M_{\pm }$ are invariant, up to a sign, under Freudenthal
duality at the horizon.

For a generic regular charge-orbit we will find the following relation
between $\mathcal{M}^H$ and the automorphism $M_+$:
\begin{equation}
\mathcal{M}^H=M_+\,\mathcal{A}\,,  \label{mmpa}
\end{equation}
where $\mathcal{A}$ is an involutive automorphism of $G$ in the stabilizer
of $\mathcal{Q}$, depending in general on $\mathcal{Q}$ and $\varphi_{flat}$%
. For $I_4<0$, both $M_+$ and $\mathcal{A}$ are anti-symplectic
outer-automorphisms of $G$, while for $I_4>0$, $\mathcal{A}\in G$ and, in
the absence of flat directions, it is the identity matrix.

Besides the interpretation in terms of $\mathcal{M}$ at the horizon, which
holds only for $M_{+}$ in specific orbits, the solution $M_{-}$ is the
symplectic metric on the $G$-orbit of $\mathcal{Q}$ \cite{LG-2} and thus it
has a mathematical relevance \textit{per se}. %
%We then consider $\mathcal{M}^{H}$ for solutions with flat directions, and
%prove a general factorization property:
%\begin{equation}
%\mathcal{M}^{H}(\mathcal{Q},\,\varphi _{flat})=\mathcal{M}_{1}^H(\mathcal{Q})%
%\mathcal{M}_{0}(\varphi _{flat})\,,  \label{m1mo}
%\end{equation}%
%where the two factor-matrices are elements of $G$ and commute. The former $%
%\mathcal{M}_{1}^H(\mathcal{Q})$ is negative definite while the latter $%
%\mathcal{M}_{0}(\varphi _{flat})$ is in the stability group of $\mathcal{Q}$%
%. As anticipated above, from this it follows that the dependence on the flat
%directions drops out once we contract $\mathcal{M}^{H}$ with a charge
%vector, so that the vector field strengths at the horizon are $\varphi
%_{flat}$-independent.
%
%The relation between the decompositions (\ref{m1mo}) and (\ref{mmpa}) is
%that $\mathcal{M}_{1}^H(\mathcal{Q})$ can be written as the product $M_+(%
%\mathcal{Q})\,\mathcal{A}_0(\mathcal{Q})$, $\mathcal{A}_0$ being an
%involutive automorphism in the stabilizer of $\mathcal{Q}$, so that $%
%\mathcal{A}(\mathcal{Q}, \varphi_{flat})$ in (\ref{mmpa}) is given by the
%product $\mathcal{A}_0(\mathcal{Q})\,\mathcal{M}_{0}(\varphi _{flat})$.

\bigskip The plan of the paper is the following.\newline
In Sect. \ref{dualsy}, we recall some basic facts about extremal black hole
solutions in extended supergravities, as well as their properties under the
global symmetry of the models. This includes a review of the Freudenthal
duality, and sets the stage for the discussion of our results.\newline
In Sec. \ref{Geom}, which focuses on the cases without flat directions, we
construct, out of a general \textit{Ansatz} involving suitable contractions
of the $K$-tensor and of the symplectic metric $\mathbb{C}_{MN}$ with a
number of charge vectors $\mathcal{Q}$, a symmetric matrix $M$ satisfying
conditions (\ref{masterequation!}). As anticipated above, restricting our
analysis to simple \textquotedblleft non-degenerate type $\mathrm{E}_{7}$%
\textquotedblright\ $U$-duality groups, treated in Subsec. \ref{Simple}, we
actually find, for $I_{4}(\mathcal{Q})>0$, two solutions: $M_{+}$ and $M_{-}$%
. The former is identified with $\mathcal{M}^{H}$, while the properties of
the latter are studied at the end of the same Section. The definition of the
matrices $M_{\pm }$ is then generalized to the $I_{4}<0$ orbit, in Sec. \ref%
{Int-2}; here general properties of $M_{\pm }$, in any regular charge-orbit $%
I_{4}\neq 0$, are discussed. \newline
In Sec. \ref{Int-3} we consider $\mathcal{N}=2$ theories, where we show that
$M_{-}$, in the BPS--orbit, is identified with the matrix $\mathcal{M}^{(F)}$%
.\newline
A general analysis, which includes the case of regular solutions with flat
directions, is finally given in Sec. \ref{Summary}, where we also summarize
the previous results. \newline
In Appendix \ref{K-Tensor}, we recall the main properties of the independent
lowest-order invariant tensors, namely $\mathbb{C}_{MN}$ (symplectic metric)
and $K_{MNPQ}$ ($K$-tensor), in the symplectic black hole charge
representation $\mathbf{R}_{\mathcal{Q}}$ of the $U$-duality groups of
\textit{symmetric} four-dimensional Maxwell-Einstein (super)gravity theories
(to which we restrict our present investigation). Appendices \ref{App-A} and %
\ref{App-B} contain details of the derivation of some results of Sec. \ref%
{Geom}, while Appendix \ref{App-C}, containing a discussion of
anti-symplectic outer-automorphisms of the U-duality algebra, concludes the
paper.

\section{\label{dualsy}Symmetry Properties of Extremal Black Holes in
Extended Supergravities}

One of the basic ingredients of the symplectic formulation of
electric-magnetic duality in $\mathcal{N}\geqslant 2$-extended supergravity
theories in four dimensions, whose bosonic Lagrangian reads (in the absence
of gauging)%
\begin{equation}
\mathcal{L}=-\frac{R}{2}+\frac{1}{2}g_{ij}\left( \varphi \right) \partial
_{\mu }\varphi ^{i}\partial ^{\mu }\varphi ^{j}+\frac{1}{4}I_{\Lambda \Sigma
}\left( \varphi \right) F_{\mu \nu }^{\Lambda }F^{\Sigma \mid \mu \nu }+%
\frac{1}{8\sqrt{-G}}R_{\Lambda \Sigma }\left( \varphi \right) \epsilon ^{\mu
\nu \rho \sigma }F_{\mu \nu }^{\Lambda }F_{\rho \sigma }^{\Sigma }\,,
\label{MEs-1}
\end{equation}%
is the $2n \times 2n $ real, negative definite, symmetric matrix $\mathcal{M}
$ \cite{BGM}:%
\begin{equation}
\mathcal{M}=\left(
\begin{array}{ccc}
\mathbb{I} &  & -R \\
&  &  \\
0 &  & \mathbb{I}%
\end{array}%
\right) \left(
\begin{array}{ccc}
I &  & 0 \\
&  &  \\
0 &  & I^{-1}%
\end{array}%
\right) \left(
\begin{array}{ccc}
\mathbb{I} &  & 0 \\
&  &  \\
-R &  & \mathbb{I}%
\end{array}%
\right) =\left(
\begin{array}{ccc}
I+RI^{-1}R &  & -RI^{-1} \\
&  &  \\
-I^{-1}R &  & I^{-1}%
\end{array}%
\right) ,  \label{M-call}
\end{equation}%
where $n$ denotes the number of Abelian vector fields, and $\mathbb{I}$
denotes the $\left( n\right) $-dimensional identity matrix. $I_{\Lambda
\Sigma }$ is the kinetic vector matrix, and $R_{\Lambda \Sigma }$ enters the
topological theta term in (\ref{MEs-1}); they are usually regarded as the
imaginary resp. real part of a complex kinetic matrix $\mathcal{N}_{\Lambda
\Sigma }$, such that (\ref{M-call}) yields $\mathcal{M}=\mathcal{M}\left[ R,I%
\right] =\mathcal{M}\left[ \text{Re}\left( \mathcal{N}\right) ,\text{Im}%
\left( \mathcal{N}\right) \right] $. Moreover, it is symplectic:
\begin{equation}
\mathcal{M}\mathbb{C}\mathcal{M}=\mathcal{M}^T\mathbb{C}\mathcal{M}=\mathbb{C%
}\,.  \label{propM1}
\end{equation}
Let us recall the main properties of this matrix which will be relevant to
our subsequent discussion.

We shall restrict our analysis to theories in which the scalar manifold is
homogeneous symmetric of the form $G/H$. The symplectic structure of the
generalized special geometry \cite{ADF-revisited,FK-N=8} of scalar fields
yields that $\mathcal{M}$ can be equivalently rewritten as%
\begin{equation}
\mathcal{M}=-\left( \mathbf{LL}^{T}\right) ^{-1}=-\mathbf{L}^{-T}\mathbf{L}%
^{-1},  \label{gen}
\end{equation}%
where $\mathbf{L}$ is an element of the $Sp\left( 2n,\,\mathbb{R}\right) $%
-valued symplectic bundle of generalized special geometry (in the symmetric
case, it is a coset representative of $G/H$ in the representation $\mathbf{R}%
_{\mathcal{Q}}$). As anticipated in the introduction, the isometry group $G$
of the scalar manifold defines the on-shell global symmetry of the theory.
Under the action of a generic isometry $g\in G$, mapping $\varphi $ into $%
\varphi ^{\prime }(\varphi )$ (to be also denoted in the following by $%
\left( g\star \varphi \right) (\varphi )=\varphi ^{\prime }(\varphi )$), $%
\mathcal{M}$ transforms as follows:
\begin{equation}
\mathcal{M}(\varphi')=
g^{-T}\,\mathcal{M}(\varphi)\,g^{-1}\,,
\label{Mtra}
\end{equation}%
the matrix $g$ representing the action of $G$ on contravariant vectors in $%
\mathbf{R}_{\mathcal{Q}}$.

The matrix $\mathcal{M}$ is an essential ingredient for writing the
equations in a manifestly symplectic-covariant way, thus making their
invariance under $U$-duality group apparent. To show this, as far as the
Maxwell equations are concerned, let us arrange the (Abelian) vector field
strengths $F^{\Lambda }$ ($\Lambda =0,1,...,n-1$; in $\mathcal{N}=2$
theories, the naught index is reserved for the graviphoton) and their
magnetic duals $G_{\Lambda }$ in a symplectic vector in the representation $%
\mathbf{R}_{\mathcal{Q}}$ of $G$:
\begin{align}
H&=(H^M)=\left( F^{\Lambda },G_{\Lambda }\right) ^{T}\,\,\,\,\,\,\left(
{}^{\ast }G_{\Lambda \mid \mu \nu }:=2\frac{\delta \mathcal{L}}{\delta
F^{\Lambda \mid \mu \nu }}\right)\,,  \label{H}
\end{align}
where ${}^*$ denotes, as usual, the Hodge-duality (which is anti-involutive
in $D=4$ spacetime: $\ast ^{2}=-Id$). This quantity satisfies the so called (%
\textit{twisted self-duality} condition) \cite{BGM}\footnote{%
Throughout this paper we use for the symplectic invariant matrix the
following form: $\mathbb{C}=\left(%
\begin{matrix}
\mathbf{0}_n & \mathbb{I}_n\cr -\mathbb{I}_n & \mathbf{0}_n%
\end{matrix}%
\right)$.}
\begin{equation}
H=\mathbb{C}\mathcal{M}~^{\ast }H,  \label{tsd}
\end{equation}
which is a symplectic-covariant relation expressing the dependence of $%
G_\Lambda$ on $F^\Lambda,\,{}^*F^\Lambda$ and the scalar fields. The Maxwell
equations are then written, in terms of $H$, as follows:
\begin{equation}
dH=0\,.
\end{equation}
Notice that consistency of the twisted self-duality condition (\ref{tsd})
with the anti-involutivity of the Hodge-operation is a direct consequence of
the symplecticity (\ref{anti-sympl}) of $\mathcal{M}$ itself. Indeed eq. (%
\ref{tsd}) can be written in the form:
\begin{align}
{}^*H=-\mathcal{S}(\varphi)\,H\,\,\,;\,\,\,\,\, \mathcal{S}( \varphi):=%
\mathbb{C}\mathcal{M}( \varphi)\,.  \label{ddd}
\end{align}
Eq. (\ref{anti-sympl}) then implies that the matrix $\mathcal{S}(\varphi)$
is actually an anti-involution:
\begin{equation}
\mathcal{S}^{2}\left( \varphi \right) =\mathbb{C}\mathcal{M}\left( \varphi
\right) \mathbb{C}\mathcal{M}\left( \varphi \right) =\mathbb{C}^{2}=-\mathbb{%
\ I}\,,
\end{equation}
representing a scalar-dependent almost-complex structure \cite{F-dual-L}
which can be defined in every generalized special geometry \cite{FK-N=8}.
For $U$-duality\footnote{%
Here $U$-duality is referred to as the \textquotedblleft
continuous\textquotedblright\ symmetries of \cite{CJ-1}. Their discrete
versions are the $U$-duality non-perturbative string theory symmetries
introduced by Hull and Townsend \cite{HT-1}.} groups $G$ \textit{of type} $%
E_{7}$ \cite{brown}, $\mathcal{S}\left( \varphi \right) \in $ $Aut(\mathbf{F}%
)\equiv G$, where $\mathbf{F}$ denotes the corresponding \textit{Freudenthal
triple system} \cite{F-dual-L}; in these theories, $\mathcal{S}$ may be
regarded as the projection onto the adjoint in the symmetric tensor product
of the symplectic representation $\mathbf{R}_{\mathcal{Q}}$ of $G$, carried
by $\mathbf{F}$ itself.

Following \cite{F-dual-L}, one defines the ``generalized scalar-dependent
Freudenthal duality''
\begin{equation}
\mathfrak{F}\,\,:\,\,\,H\,\,\longrightarrow\,\,\, \mathfrak{F}(H):=-\mathcal{%
S}(\varphi)\,H\,,  \label{fd0}
\end{equation}
whose general features are discussed in the same paper. By the above
properties of the matrix $\mathcal{S}(\varphi)$, $\mathfrak{F}$ is
anti-involutive: $\mathfrak{F}^2=-Id$. The compatibility of the two
anti-involutive structures, defined by $\mathfrak{F}$ and the
Hodge-operation ${}^*$, directly follows from (\ref{tsd}) and the
anti-involutivity of $S(\varphi)$ \cite{F-dual-L}:
\begin{equation}
H=-\mathfrak{F}\left( ^{\ast }H\right) =-^{\ast }\mathfrak{F}\left(
H\right)\,.  \label{tsdd}
\end{equation}

The matrix $\mathcal{M}$ plays an important role in the study of the
properties of black hole solutions to ungauged extended supergravities, in
relation to the $U$-duality group of the model. In the background of a
static, spherically symmetric, asymptotically flat, dyonic extremal black
hole ($\mathbf{\tau }:=-1/r$)%
\begin{equation}
ds^{2}=-e^{2U\left( \mathbf{\tau }\right) }dt^{2}+e^{-2U\left( \mathbf{\tau }%
\right) }\left[ \frac{d\mathbf{\tau }^{2}}{\mathbf{\tau }^{4}}+\frac{1}{%
\mathbf{\tau }^{2}}\left( d\theta ^{2}+\sin \theta d\psi ^{2}\right) \right]
,  \label{r-2}
\end{equation}%
one can introduce the symplectic vector $\mathcal{Q}=\{p^{\Lambda
},q_{\Lambda }\}$ of asymptotic magnetic and electric fluxes of $H$ as
follows:%
\begin{equation*}
\mathcal{Q}=\frac{1}{4\pi }\,\int_{S^{2}}H\,\,\Leftrightarrow
\,\,\,\,\,\,p^{\Lambda }=\frac{1}{4\pi }\int_{S^{2}}F^{\Lambda },~q_{\Lambda
}=\frac{1}{4\pi }\int_{S^{2}}G_{\Lambda }\,,
\end{equation*}%
$S^{2}$ being any sphere of radius $r$. The spherical symmetry requires the
scalar fields to depend on $\tau $ (or equivalently $r$) only. The action of
a generic global symmetry transformation $g$ in $G$ maps a black hole in
this class, described by scalar fields $\varphi (\tau )=(\varphi ^{i}(\tau
)) $ and a charge vector $\mathcal{Q}$, into a solution of the same kind
with a charge vector $\mathcal{Q}^{\prime }=g\,\mathcal{Q}$ and scalar
fields $\varphi ^{\prime }(\tau )=g\star \varphi (\tau )$:
\begin{equation}
g\in G\,\,\,:\,\,\,\,\,[\varphi (\tau ),\,\mathcal{Q}]\,\,\longrightarrow
\,\,\,\,[g\star \varphi (\tau ),\,g\,\mathcal{Q}]\,.  \label{utra}
\end{equation}%
The generalized Freudenthal duality in (\ref{fd0}) induces a
scalar-dependent transformation on the electric-magnetic charges
\begin{equation*}
\mathcal{Q}\longrightarrow \,\,\mathfrak{F}(\mathcal{Q})=\mathfrak{F}\left(
\frac{1}{4\pi }\,\int_{S^{2}}H\right) :=\frac{1}{4\pi }\,\int_{S^{2}}%
\mathfrak{F}(H)=-\mathcal{S}(\varphi )\,\mathcal{Q}\,.
\end{equation*}%
The action of $\mathfrak{F}$ on $\mathcal{Q}$ represents the
\textquotedblleft non-critical", scalar-dependent generalization of the
so-called \textit{Freudenthal duality} \cite{Duff-FD}, defined first in \cite%
{FMY-FD-1}. Condition (\ref{tsd}) then implies that:
\begin{equation}
\mathfrak{F}(\mathcal{Q})=\frac{1}{4\pi }\,\int_{S^{2}}{}^{\ast }H\,.
\end{equation}%
The Abelian $2$-form field strengths $H$, in the background (\ref{r-2}), can
be written, using the matrix $\mathcal{M}$, in the following $\mathrm{Sp}(2n,%
\mathbb{R})$-covariant form (\textit{cfr. e.g.} \cite%
{Denef-1,ADFT-rev,ADFT-Small-1})%
\begin{eqnarray}
H\left( \varphi ,U,\mathcal{Q}\right) &=&e^{2U}\mathbb{C}\mathcal{M}\left(
\varphi \right) \mathcal{Q}dt\wedge d\mathbf{\tau }+\mathcal{Q}\sin \theta
d\theta \wedge d\psi  \notag \\
&=&-e^{2U}\mathfrak{F}\left( \mathcal{Q}\right) dt\wedge d\mathbf{\tau }+%
\mathcal{Q}\sin \theta d\theta \wedge d\psi \,,  \label{fs-1}
\end{eqnarray}%
thus implying that (recall (\ref{tsd}))%
\begin{equation*}
^{\ast }H(\varphi ,U,\mathcal{Q})=\mathfrak{F}(H(\varphi ,U,\mathcal{Q}%
))=e^{2U}\mathcal{Q}dt\wedge d\tau +\mathfrak{F}(\mathcal{Q})\sin \theta
d\theta \wedge d\psi =H(\varphi ,U,\mathfrak{F}(\mathcal{Q})),
\end{equation*}%
consistently with (\ref{tsdd}). Note that the dependence of $H$ on the
scalars is completely encoded in $\mathcal{M}\left( \varphi \right) $, or,
equivalently, in the \textquotedblleft non-critical" Freudenthal duality $%
\mathfrak{F}$ (\ref{fd0}).

$\mathcal{M}$ also defines the (positive definite) effective black hole
potential (\ref{V_BH}), such that $\mathfrak{F}$ (\ref{fd0}) can
equivalently be defined as%
\begin{equation}
\mathfrak{F}:\mathcal{Q\rightarrow }\mathfrak{F}\left( \mathcal{Q}\right) :=%
\mathbb{C}\frac{\partial V_{BH}}{\partial \mathcal{Q}}.  \label{FHdef}
\end{equation}%
The potential $V_{BH}$ (\ref{V_BH}) governs the radial evolution of the
scalar fields $\varphi (\mathbf{\tau })$ as well as of the warp factor $%
U\left( \mathbf{\tau }\right) $:%
\begin{equation*}
\frac{d^{2}U}{d\mathbf{\tau }^{2}}=e^{2U}V_{BH}\,\,\,;\,\,\,\,\frac{%
d^{2}\varphi ^{i}}{d\mathbf{\tau }^{2}}=g^{ij}e^{2U}\frac{\partial V_{BH}}{%
\partial \varphi ^{j}}.
\end{equation*}%
By virtue of (\ref{Mtra}), $V_{BH}(\varphi ,\mathcal{Q})$ is invariant under
a $U$-duality transformation (\ref{utra}):
\begin{equation}
\forall g\in G\,\,:\,\,\,\,V_{BH}(g\star \varphi ,\,g\,\mathcal{Q}%
)=V_{BH}(\varphi ,\,\mathcal{Q})\,.  \label{Vinv}
\end{equation}%
At the event horizon of an extremal black hole\footnote{%
Here we shall restrict to the so called \textquotedblleft
large\textquotedblright , \textit{i.e.} regular, extremal black holes,
namely to solutions whose singularity is hidden inside an event horizon with
finite area $A_{H}$. These solutions are characterized by the property $%
A_{H}=4\pi \,\sqrt{|I_{4}|}\neq 0$, see Eqs. (\ref{SS_BH}) and (\ref{S_BH})
below, \textit{i.e.} that the quartic invariant $I_{4}$, defined below,
computed on their electric and magnetic charges, is non-vanishing.} ($%
\mathbf{\tau }\rightarrow -\infty $), the \textit{attractor mechanism} \cite%
{AM,FGK} implies that, regardless of the initial (asymptotic) conditions,
the scalar fields evolve towards values $\varphi _{H}^{i}(\mathcal{Q})$
which only depend, up to \textit{flat} directions \cite{FM-Moduli-Spaces},
on the quantized charges:
\begin{equation}
\text{lim}_{\mathbf{\tau }\rightarrow -\infty }\varphi ^{i}=\varphi
_{H}^{i}\left( \mathcal{Q}\right) \,.  \label{AM}
\end{equation}%
The fixed point $\varphi _{H}$ corresponds to the minimum of $V_{BH}$:
\begin{equation}
\left. \frac{\partial V_{BH}}{\partial \varphi ^{i}}\right\vert _{\varphi
=\varphi _{H}}=0\,.
\end{equation}%
Flat directions, generically denoted by $\varphi _{flat}$, are scalar fields
which $V_{BH}$ does not depend on, and thus they are not fixed by the above
extremality condition (\textit{at least} at Einsteinian level \cite%
{FM-Moduli-Spaces}). These directions in symmetric supergravities span a
symmetric submanifold of the scalar manifold of the form \cite%
{FM-Moduli-Spaces}:
\begin{equation}
\varphi _{flat}\,\in \,\frac{G_{0}}{H_{0}}\,\subset \,\frac{G}{H}\,,
\end{equation}%
where $G_{0}$ is the \emph{stabilizer} in $G$ of the charge vector $\mathcal{%
Q}$ and $H_{0}$ its maximal compact subgroup.

Excluding, for the time being, the existence of $\varphi _{flat}$, which
shall be dealt with separately, the $U$-duality invariance (\ref{Vinv}) of $%
V_{BH}$ implies that
\begin{equation}
\varphi _{H}(g\,\mathcal{Q})=g\star \varphi _{H}(\mathcal{Q})\,.
\label{phiHQ}
\end{equation}%
In the near-horizon limit also $\mathcal{M}$, computed on the solution, will
evolve towards a matrix $\mathcal{M}^{H}$, defined as
\begin{equation}
\mathcal{M}^{H}:=\lim_{\mathbf{\tau }\rightarrow -\infty }\mathcal{M}\left(
\varphi \left( \mathbf{\tau }\right) \right) =\mathcal{M}(\varphi _{H}^{i})=%
\mathcal{M}^{H}(\mathcal{Q})\,.  \label{MH-def}
\end{equation}%
We now introduce a set of \emph{dual} charges $\tilde{\mathcal{Q}}=(\tilde{%
\mathcal{Q}}^{M})$ defined as:
\begin{equation}
\mathcal{\tilde{Q}}:=\text{lim}_{\mathbf{\tau }\rightarrow -\infty }%
\mathfrak{F}\left( \mathcal{Q}\right) =-\mathbb{C}\mathcal{M}^{H}\mathcal{Q}%
\,,  \label{tildeQM}
\end{equation}%
which defines a \textquotedblleft critical\textquotedblright\ Freudenthal
duality \cite{Duff-FD}:
\begin{equation}
\mathfrak{F}_{H}\left( \mathcal{Q}\right) :=\mathcal{\tilde{Q}}=-\mathcal{S}%
^{H}\,\mathcal{Q}\,,
\end{equation}%
where $\mathcal{S}^{H}:=\mathbb{C}\mathcal{M}^{H}$. It can be shown \cite%
{FMY-FD-1} that
\begin{equation}
\mathcal{\tilde{Q}}=\frac{1}{\pi }\mathbb{C}\frac{\partial S_{BH}}{\partial
\mathcal{Q}}\,,  \label{FD-def-H}
\end{equation}%
where $S_{BH}$ denotes the Bekenstein-Hawking entropy \cite{Bek-Haw} of the
extremal black hole (\ref{r-2}), given by
\begin{equation}
S_{BH}=\frac{A_{H}}{4}=-\frac{\pi }{2}\mathcal{M}_{MN}^{H}\mathcal{Q}^{M}%
\mathcal{Q}^{N}\,.  \label{SS_BH}
\end{equation}%
Note that (\ref{SS_BH}) implies that $\mathcal{M}^{H}$ is homogeneous of
degree zero in the charges.

For $U$-duality groups \textit{of type} $E_{7}$ \cite{brown}, $\mathcal{%
\tilde{Q}}$ can also be written as \cite{Duff-FD,FMY-FD-1}%
\begin{equation}
\mathcal{\tilde{Q}}_{M}=\epsilon \,\frac{2}{\sqrt{\left\vert
I_{4}\right\vert }}K_{MNPQ}\mathcal{Q}^{N}\mathcal{Q}^{P}\mathcal{Q}^{Q}\,,
\label{Qtilde}
\end{equation}%
where $\epsilon =\pm 1$, the index $M$ was lowered by means of $\mathbb{C}$,
$\mathcal{\tilde{Q}}_{M}=\mathbb{C}_{NM}\,\tilde{\mathcal{Q}}^{N}$, and $%
K_{MNPQ}$ is the so-called $K$-tensor, which is the invariant tensor in the
4-fold symmetric product of the representation $\mathbf{R}_{\mathcal{Q}}$,
whose properties are summarized in Appendix \ref{K-Tensor}. In terms of it,
one can express the invariant quartic homogeneous polynomial $I_{4}$ in the
charges $\mathcal{Q}$ as:%
\begin{equation}
I_{4}:=K_{MNPQ}\mathcal{Q}^{M}\mathcal{Q}^{N}\mathcal{Q}^{P}\mathcal{Q}%
^{Q}=\epsilon \,\left\vert I_{4}\right\vert ,  \label{I4-def}
\end{equation}%
thus implying that the Bekenstein-Hawking entropy $S_{BH}$ (\ref{SS_BH}) can
be written as%
\begin{equation}
S_{BH}=\pi \sqrt{\left\vert I_{4}\right\vert }\,.  \label{S_BH}
\end{equation}%
The above expression of the entropy is manifestly invariant under a
\textquotedblleft critical\textquotedblright\ (as well as \textquotedblleft
non-critical\textquotedblright ) Freudenthal duality, since the latter
amounts to acting on the charge vector by means of $\mathcal{S}^{H}$ (or $%
\mathcal{S}$ in the \textquotedblleft non-critical\textquotedblright\ case),
which is an element of $G$.

Using Eqs. (\ref{tildeQM}) and (\ref{Qtilde}), we find that the
charge-dependent matrix $\mathcal{M}^{H}$ satisfies the following
distinctive property:
\begin{equation}
\mathcal{M}^{H}\mathcal{Q}=-\frac{\epsilon }{2\sqrt{\left\vert
I_{4}\right\vert }}\frac{\partial I_{4}}{\partial \mathcal{Q}}=-2\frac{%
\epsilon }{\sqrt{\left\vert I_{4}\right\vert }}K_{MNPQ}\mathcal{Q}^{M}%
\mathcal{Q}^{N}\mathcal{Q}^{P}\mathcal{Q}^{Q}\,,  \label{propM2}
\end{equation}%
which clearly implies $\mathcal{Q}^{T}\mathcal{M}^{H}\,\mathcal{Q}=-2\,\sqrt{%
|I_{4}|}=-2\,S_{BH}/\pi $, \textit{i.e.} Eq. (\ref{SS_BH}).

From the above discussion we conclude that, at the event horizon of the
extremal black hole, the symplectic field strength vector%
\begin{equation}
H_{H}:=\text{lim}_{\mathbf{\tau }\rightarrow -\infty }H\left( \varphi \left(
\mathbf{\tau }\right) \right)  \label{H_H}
\end{equation}%
reads%
\begin{eqnarray}
H_{H} &=&e^{2U_{H}}\mathbb{C}\mathcal{M}^{H}\mathcal{Q}dt\wedge d\mathbf{%
\tau }+\mathcal{Q}\sin \theta d\theta \wedge d\psi =  \notag \\
&=&-e^{2U_{H}}\mathcal{\tilde{Q}}dt\wedge d\mathbf{\tau }+\mathcal{Q}\sin
\theta d\theta \wedge d\psi =-\mathfrak{F}_{H}\left( ^{\ast }H_{H}\right) ,
\label{deff-1}
\end{eqnarray}%
where $U_{H}$ is the leading order contribution in $\mathbf{\tau }$ of the
near-horizon limit of $U(\mathbf{\tau })$. From Eq. (\ref{propM2}) it
follows that, in the presence of flat directions, although $\mathcal{M}^{H}$
in general depends on them, the fields strengths $H_{H}$ near the horizon do
not.

The expression of the matrix $\mathcal{M}$ evaluated on the radial flow of
the scalar fields in a black hole solution, can be rather complicated due to
the highly non-linear dependence that $\mathcal{M}$ can have on the scalars $%
\varphi $ (in generic $d$-geometries, for instance, the expression of the
real symmetric matrices\ $I_{\Lambda \Sigma }$ and $R_{\Lambda \Sigma }$ can
be found \textit{e.g.} in Sec. 2 of \cite{CFM-1}; see also App. A of \cite%
{Perz}). One would however expect, by virtue of the attractor mechanism, the
near-horizon behavior of the matrix $\mathcal{M}$ to simplify considerably,
since all the physical properties of the solution, in this limiting region,
only depend on the quantized charges $p^{\Lambda },\,q_{\Lambda }$.
Characterizing this behavior is the main motivation of our investigation.

As previously pointed out, in the absence of \textit{flat} directions, when
\textit{all} the scalars $\varphi $'s are stabilized to a (purely) $\mathcal{%
Q}$-dependent value $\varphi _{H}\left( \mathcal{Q}\right) $ (\ref{AM}) at
the horizon, by the attractor mechanism, also the limiting value $\mathcal{M}%
^{H}$ of $\mathcal{M}$ is a function of $\mathcal{Q}$ only, see eq. (\ref%
{MH-def}). Consequently, the action of an element $g$ of the $U$-duality
group $G$ on the solution, which maps the initial charge vector $\mathcal{Q}$
into $\mathcal{Q}^{\prime }=g\,\mathcal{Q}$, induces a linear transformation
on $\mathcal{M}^{H}$, as a result of Eqs. (\ref{Mtra}),(\ref{MH-def}) and (%
\ref{phiHQ}):
\begin{equation}
\mathcal{M}^{H}(g\,\mathcal{Q})=\mathcal{M}(\varphi _{H}(g\,\mathcal{Q}))=%
\mathcal{M}(g^{\mathbf{\star }}\varphi _{H}(\mathcal{Q}))=g^{-T}\mathcal{M}%
(\varphi _{H}(\mathcal{Q}))g^{-1}=g^{-T}\mathcal{M}^{H}(\mathcal{Q})g^{-1}.
\label{Mflow}
\end{equation}%
The above transformation property hints at some intrinsic group-theoretical
characterization of $\mathcal{M}^{H}$ since any symmetric $\mathrm{Sp}(2n,%
\mathbb{R})$-covariant matrix $M(\mathcal{Q})_{MN}$, built out of the charge
vector $\mathcal{Q}$ and of $G$-invariant tensors, transforms under $G$ as $%
\mathcal{M}^{H}$ in (\ref{Mflow}). Certainly an $Sp(2n_{V}+2,\mathbb{R})$%
-covariant, symmetric matrix $M(\mathcal{Q})$, only built out of $\mathcal{Q}
$ and of $G$-invariant tensors in products of the representation $\mathbf{R}%
_{\mathcal{Q}}$, satisfies the above transformation property. These $G$%
-invariant tensors in products of the representation $\mathbf{R}_{\mathcal{Q}%
}$ include the symplectic-invariant metric $\mathbb{C}_{MN}$ and the rank-$4$
completely symmetric invariant $K$-tensor $K_{MNPQ}$ (\textit{cfr.} Appendix %
\ref{K-Tensor}). In the next sections we address the problem of expressing $%
\mathcal{M}^{H}$ in terms of a matrix $M(\mathcal{Q})$ of this kind,
restricting ourselves to $D=4$ Maxwell-Einstein (super)gravity theories
whose scalar manifold is a \textit{symmetric} space $G/H$ (which correspond
to characterizing $G$ as a group \textit{of type} $E_{7}$ \cite{brown}). We
find a simple identification for the $I_{4}>0$ orbits in the absence of flat
directions. For a generic orbit, on the other hand, we will be able to
identify $\mathcal{M}^{H}$ with a charge-dependent $G$-covariant matrix,
modulo the multiplication by an involutive automorphism of $G$ in the
stabilizer of $\mathcal{Q}$.

\section{\label{Geom}The $M$-Matrix and $\mathcal{M}^{H}$ \textit{without}
Flat Directions}

In the present section we focus on a class of four-dimensional
Maxwell-Einstein (super)gravity theories with \textit{symmetric} scalar
manifolds $G/H$. We look for a matrix $M(\mathcal{Q})_{MN}$ constructed out
of $\mathcal{Q}$ and of the $G$-invariant structures $K_{MNPQ}$, $\mathbb{C}%
_{MN}$, which satisfy the distinctive properties (\ref{propM1}), (\ref%
{propM2}) of $\mathcal{M}^{H}(\mathcal{Q})_{MN}$. These conditions turn out
to be rather restrictive and, starting from a general \textit{Ansatz} for $M(%
\mathcal{Q})$, we were able to find a solution only in the $I_{4}(\mathcal{Q}%
)>0$ orbit. We were able instead, for a generic orbit of $\mathcal{Q}$, to
find solutions to the equations
\begin{equation}
M\mathbb{C}M=\epsilon \,\mathbb{C}\,\,\,\,;\,\,\,\,\,\,M\mathcal{Q}=-\frac{%
\epsilon }{2\sqrt{\left\vert I_{4}\right\vert }}\frac{\partial I_{4}}{%
\partial \mathcal{Q}}\,,  \label{meq2}
\end{equation}%
where $I_{4}(\mathcal{Q})=\epsilon \,|I_{4}(\mathcal{Q})|$. Notice the
difference between the first of the above equations and (\ref{propM1}), in
the presence of the sign $\epsilon $ of $I_{4}$ on the right hand side of
the former. In fact, for $\epsilon =-1$, the first of eq.s (\ref{meq2})
defines an \emph{anti-symplectic} symmetric matrix instead of a symplectic
one. For each regular orbit ($I_{4}>0,\,I_{4}<0$), we find two distinct
solutions $M_{\pm }$ with different properties.

In the present investigation we shall only consider simple $U$-duality
groups $G$ \emph{non-degenerate of type $\mathrm{E}_{7}$} \cite%
{brown,FKM-Deg-E7} (see footnote 7 below), leaving the treatment of the
other cases to future work. In the absence of flat directions and for $%
I_{4}>0$ ($\epsilon =+1$), we can identify one of the two matrices ($M_{+}$)
with $\mathcal{M}_{MN}^{H}$. Thus, even if the definiton of $M_{\pm }$ is
general, the identification $\mathcal{M}_{H}=M_{+}$ turns out to hold only
for (\textit{cfr.} \textit{e.g.} \cite{FM-Moduli-Spaces}):

\begin{enumerate}
\item ($1/2$-)BPS attractors in all $\mathcal{N}=2$ simple \emph{%
non-degenerate}-\emph{type $\mathrm{E}_{7}$} symmetric models (we exclude
from our analysis the \emph{minimal coupling} ones);

\item non-BPS $Z_{H}=0$ attractors in $STU$ model with $I_{4}>0$.
\end{enumerate}

As mentioned in the Introduction, in the most general case, $\mathcal{M}^H$
coincides with $M_+$ \emph{modulo} multiplication by a transformation $%
\mathcal{A}$ in the little group of $\mathcal{Q}$. For $I_4<0$ ($\epsilon
=-1 $), $\mathcal{A}$ is non-trivial also in the absence of flat directions,
as in the case of the $\mathcal{N}=2$ $T^3$-model, since $M_+$ is
antisymplectic as opposed to $\mathcal{M}^H$.

\emph{Non-degenerate} $U$-duality groups $G$ \textquotedblleft of type $%
E_{7} $" \cite{brown,FKM-Deg-E7} will be considered in Subsec.s \ref{Simple}%
, \ref{Int-2} and \ref{Int-3}. Here we first construct the solutions $M_{\pm
}$ for $I_{4}>0$, discuss their geometric properties and the relation of one
of them to $\mathcal{M}^{H}$. Then we move to the definition of $M_{\pm }$
in the $I_{4}<0$ case, generalizing some of their properties to all regular
orbits. %
%The particular case of \textit{minimal coupling }of Abelian vector
%multiplets to $\mathcal{N}=2$ supergravity, in which $K_{MNPQ}$ is \textit{%
%reducible} (corresponding to \textit{degenerate} groups \textquotedblleft of
%type $E_{7}$" \cite{FKM-Deg-E7}), will be considered in Subsec. \ref{MC}.

\subsection{\label{Simple}The $I_4>0$ Case and $\mathcal{M}^H$}

We start by considering the orbit $I_4>0$ (i.e. $\epsilon=+1$) of $\mathcal{Q%
}$ and we look for a $G$ -covariant symmetric matrix $M(\mathcal{Q})$,
solution to the equations (\ref{masterequation!}):
\begin{eqnarray}
M_{MN}M_{PQ}\mathbb{C}^{NP} &=&\mathbb{C}_{MQ}\,;  \label{SymplCond} \\
M_{MN}\mathcal{Q}^{N} &=&-\frac{1}{2\sqrt{\left\vert I_{4}\right\vert }}%
\frac{\partial I_{4}}{\partial \mathcal{Q}^{M}}=-\mathcal{\tilde{Q}}_{M}.
\label{M0QQ}
\end{eqnarray}%
We use for $M$ the following general \textit{Ansatz } \textit{\ } ($A$, $B$,
$C\in \mathbb{R}$):%
\begin{equation}
M_{MN}(\mathcal{Q})=\frac{A}{|I_{4}|^{3/2}}\,K_{M}K_{N}+\frac{B}{%
|I_{4}|^{1/2}}K_{MN}+\frac{C\,}{|I_{4}|^{1/2}}K_{MB_{1}B_{2}}K_{NB_{3}B_{4}}%
\mathbb{C}^{B_{1}B_{3}}\mathbb{C}^{B_{2}B_{4}},  \label{Ans-1}
\end{equation}%
where:%
\begin{equation}
K_{MNP}:=K_{MNPQ}\mathcal{Q}^{Q},~K_{MN}:=K_{MNPQ}\mathcal{Q}^{P}\mathcal{Q}%
^{Q},~K_{M}:=K_{MNPQ}\mathcal{Q}^{N}\mathcal{Q}^{P}\mathcal{Q}^{Q}.
\end{equation}%
The derivations below strongly rely on the properties of the $K$-tensor, for
simple $G$, discussed in Appendix \ref{K-Tensor}. By recalling (\ref{Qtilde}%
) \cite{Duff-FD,FMY-FD-1}, it holds that%
\begin{equation}
K_{M}=\frac{1}{2}\epsilon \left\vert I_{4}\right\vert ^{1/2}\mathcal{\tilde{Q%
}}_{M},
\end{equation}%
such that (\ref{Ans-1}) can be rewritten as%
\begin{equation}
M_{MN}(\mathcal{Q})=\frac{A}{4|I_{4}|^{1/2}}\,\mathcal{\tilde{Q}}_{M}%
\mathcal{\tilde{Q}}_{N}+\frac{B}{|I_{4}|^{1/2}}K_{MN}+\frac{C}{|I_{4}|^{1/2}}%
\,K_{MB_{1}B_{2}}K_{NB_{3}B_{4}}\mathbb{C}^{B_{1}B_{3}}\mathbb{C}%
^{B_{2}B_{4}}.  \label{Ans-2}
\end{equation}

By exploiting the identity\footnote{%
As discussed in \cite{Exc-Reds} and in \cite{FKM-Deg-E7}, this is a
consequence of a general identity involving the quantity $%
K_{MNPA_{1}}K_{PQRA_{2}}\mathbb{C}^{A_{1}A_{2}}$, given by (5.16) of \cite%
{Exc-Reds}.}%
\begin{equation}
K_{MA_{1}A_{2}}K_{PA_{3}A_{4}}\mathbb{C}^{A_{1}A_{3}}\mathbb{C}%
^{A_{2}A_{4}}=-\frac{1}{6\tau }\left[ (2\tau -1)K_{MP}+\frac{1}{12}\left(
\tau -1\right) \,\,\mathbb{C}_{A_{1}(M}\mathbb{C}_{P)A_{2}}\mathcal{Q}%
^{A_{1}}\mathcal{Q}^{A_{2}}\right] \,,  \label{KKCC-1}
\end{equation}%
where $\tau$ is defined in eq. (\ref{tau}), the \textit{Ansatz} (\ref{Ans-1}%
) (or, equivalently (\ref{Ans-2})) can be further simplified as%
\begin{eqnarray}
M_{MN}(\mathcal{Q}) &=&\frac{A}{|I_{4}|^{3/2}}\,K_{M}K_{N}+\frac{1}{%
|I_{4}|^{1/2}}\left( B+\frac{\left( 1-2\tau \right) }{6\tau }C\right) K_{MN}+%
\frac{C}{72|I_{4}|^{1/2}}\frac{\left( \tau -1\right) }{\tau }\mathcal{Q}_{M}%
\mathcal{Q}_{N}  \notag \\
&&  \label{Ans-3} \\
&=&\frac{A}{4|I_{4}|^{1/2}}\,\mathcal{\tilde{Q}}_{M}\mathcal{\tilde{Q}}_{N}+%
\frac{1}{|I_{4}|^{1/2}}\left( B+\frac{\left( 1-2\tau \right) }{6\tau }%
C\right) K_{MN}+\frac{C}{72|I_{4}|^{1/2}}\frac{\left( \tau -1\right) }{\tau }%
\mathcal{Q}_{M}\mathcal{Q}_{N}.  \notag \\
&&  \label{Ans-4}
\end{eqnarray}

In App. \ref{App-A}, the real coefficients $A,\,B$ and $C$ in (\ref{Ans-1})
and (\ref{Ans-2}) are determined by exploiting the properties (\ref{meq2}).

It should be remarked that a term proportional to $\mathcal{Q}_{(M}\mathcal{%
\tilde{Q}}_{N)}$ cannot occur in (\ref{Ans-2}) (or, equivalently, in (\ref%
{Ans-4})), because it is not consistent with (\ref{M0QQ}) \cite{FMY-FD-1}.

A consistent solution to (\ref{SymplCond})-(\ref{M0QQ}) within the \textit{%
Ansatz} (\ref{Ans-1}) can be found only for $\epsilon =+1\Leftrightarrow
I_{4}>0$, and it reads%
\begin{equation}
A_{\pm }=-2\mp \,6\,,\quad B_{\pm }=\frac{6\left( 1-2\tau \mp \tau \right) }{%
(\tau -1)}\,,\quad C_{\pm }=-\frac{36\tau \left( 1\pm 1\right) }{\,(\tau -1)}%
\,.  \label{ABC-1}
\end{equation}%
The splitting into \textquotedblleft $\pm $\textquotedblright\ branches
generally corresponds to two independent expressions, namely $M_{+}$ and $%
M_{-}$, in terms of suitable contractions of the $K$-tensor itself and of
the symplectic metric $\mathbb{C}_{MN}$ with charge vectors $\mathcal{Q}$'s;
note that $M_{-}$ lacks the term proportional to $\mathcal{Q}_{M}\mathcal{Q}%
_{N}$, because $C_{-}=0$. From eq.s (\ref{Ans-4}), (\ref{ABC-1}), see
Appendix \ref{withoutS}, we can write the two solutions in a universal form:
\begin{equation}
M_{\pm |MN}(\mathcal{Q})=-\frac{2\pm 6}{\left\vert I_{4}\right\vert ^{3/2}}%
\,K_{M}K_{N}\pm \frac{6}{\left\vert I_{4}\right\vert ^{1/2}}\,K_{MN}-\frac{%
1\pm 1}{2\left\vert I_{4}\right\vert ^{1/2}}\mathcal{Q}_{M}\mathcal{Q}_{N},
\label{finalMpm}
\end{equation}

This \textquotedblleft $\pm $\textquotedblright\ ambiguity can be removed
when considering the relation to the negative-definite matrix $\mathcal{M}^H$%
. Indeed $M_{-}(\mathcal{Q})$ always has (\textit{at least}) a positive
eigenvalue and thus can never be identified with $\mathcal{M}^H$. This
result is illustrated in App. \ref{App-B} by a direct computation in the $%
STU $ model (and its rank-2 ($ST^{2}$) and rank-1 ($T^{3}$)
\textquotedblleft degenerations" determine the corresponding symmetric
models), and thus holds at least in all rank-$3$ symmetric models of which
the $STU$ one is a universal sector. This check allows one to conclude that
only the \textquotedblleft $+ $" branch can be consistent with the
properties required for the matrix $\mathcal{M}$ (at the horizon).

Using (\ref{finalMpm}), direct computations in the STU model and its
contractions (e.g. the $T^{3}$ model) suggests the following identification
(recall $I_{4}>0$)
\begin{eqnarray}
\mathcal{M}_{MN}^{H}(\mathcal{Q}) &=&M_{+\mid MN}(\mathcal{Q})=-\frac{1}{%
\sqrt{I_{4}}}\left( \frac{8}{I_{4}}\,K_{M}K_{N}-6K_{MN}+\mathcal{Q}_{M}%
\mathcal{Q}_{N}\right)  \notag \\
&=&-\frac{1}{\sqrt{I_{4}}}\left( 2\,\mathcal{\tilde{Q}}_{M}\mathcal{\tilde{Q}%
}_{N}-6K_{MN}+\mathcal{Q}_{M}\mathcal{Q}_{N}\right) \,,  \label{Res-5}
\end{eqnarray}%
which, as far as the STU model is concerned, holds for both the BPS and
non-BPS orbits $I_{4}>0$.

Let us now show that, once proven for the STU model (and its
\textquotedblleft degenerations" ST$^{2}$ and T$^{3}$ models), the above
identification holds for the BPS solutions to \textit{any} symmetric $%
\mathcal{N}=2$ theory of which the STU model (or its \textquotedblleft
degenerations") is a consistent truncation\footnote{%
This class of models have the feature that $G$ is of \emph{type $\mathrm{E}%
_{7}$} and does include the \emph{minimal-coupling} models with special K%
\"{a}hler manifold $\frac{\mathrm{SU}(1,n)}{\mathrm{U}(n)}$ only as a \emph{%
degenerate} \cite{FKM-Deg-E7} instance, which we shall not be dealing with
in this paper. An other class of \emph{degenerate}-\emph{type $\mathrm{E}%
_{7} $} models are the $\mathcal{N}=3$ supergravities, with scalar manifold $%
\frac{\mathrm{SU}(3,n)}{\mathrm{S}[\mathrm{U}(3)\times \mathrm{U}(n)]}$,
which will be dealt with elsewhere.}. These comprise all the theories
originating from dimensional reduction from $D=5$ and include the
\textquotedblleft magical\textquotedblright\ ones \cite{GST}. The
corresponding symmetric special K\"{a}hler manifold $G/H$ has the isotropy
group $H$ of the form $H=\mathrm{U}(1)\times \mathcal{H}_{0}$, where $%
\mathcal{H}_{0}$ is the compact real form of the duality group in $D=5$ and
is also isomorphic in $G$ to the stability group $G_{0}$ of a charge vector $%
\mathcal{Q}$ in the BPS orbit. This group, being compact, coincides with its
maximal compact subgroup $H_{0}$, so that $\mathcal{H}_{0}$ and $H_{0}$ are
isomorphic in $G$. With respect to $\mathcal{H}_{0}$ (or, equivalently $%
H_{0} $) the representation $\mathbf{R}_{\mathcal{Q}}$ branches as follows:
\begin{equation}
\mathbf{R}_{\mathcal{Q}}\,\,\,\overset{\mathcal{H}_{0}}{\longrightarrow }%
\,\,\,\,\mathbf{1}+\mathbf{R}+\bar{\mathbf{1}}+\bar{\mathbf{R}}\,,
\label{RqR}
\end{equation}%
where $\mathbf{R}$ is, for the \textquotedblleft magical\textquotedblright\
theories, an irreducible representation. We can choose a representative $%
\mathcal{Q}$ of the BPS orbit whose stabilizer $H_{0}$ coincides with the
isotropy group $\mathcal{H}_{0}$ of the manifold. The components of the
vector $\mathcal{Q}$ correspond to the singlets $\mathbf{1}+\bar{\mathbf{1}}$
in (\ref{RqR}). The charges in the STU truncation comprise these two
singlets and six components in the representation $\mathbf{R}+\bar{\mathbf{R}%
}$, defining the \emph{normal form} of a generic element of $\mathbf{R}$
with respect to the action of $\mathcal{H}_{0}$. Both the two matrices $%
\mathcal{M}^{H}(\mathcal{Q})$ and $M_{+}(\mathcal{Q})$ commute with $%
\mathcal{H}_{0}$:
\begin{equation*}
\forall h\in \mathcal{H}_{0}\,\,\,:\,\,\,%
\begin{cases}
h\,\mathcal{M}^{H}(\mathcal{Q})\,h^{T}=\mathcal{M}^{H}(h\,\mathcal{Q})=%
\mathcal{M}^{H}(\mathcal{Q})\cr h\,M_{+}(\mathcal{Q})\,h^{T}=M_{+}(h\,%
\mathcal{Q})=M_{+}(\mathcal{Q})%
\end{cases}%
\,\,\Leftrightarrow \,\,\,\,%
\begin{cases}
\lbrack h,\,\mathcal{M}^{H}(\mathcal{Q})]=0\cr [h,\,M_+(\mathcal{Q})]=0%
\end{cases}%
\,,
\end{equation*}%
where we have used the properties that the symplectic duality action of $%
\mathcal{H}_{0}$ is represented by orthogonal matrices and that $\mathcal{H}%
_{0}$ is the stabilizer of $\mathcal{Q}$. If $\mathbf{R}$ is irreducible, by
Schur's lemma, $\mathcal{M}^{H}(\mathcal{Q})$ and $M_{+}(\mathcal{Q})$ are
both proportional to the identity on $\mathbf{R}$ and thus, since they
coincide on the STU model, which comprise charges in $\mathbf{R}$, they do
coincide on the whole $\mathbf{R}_{\mathcal{Q}}$.

As for the infinite series of models with special K\"{a}hler manifold $\frac{%
\mathrm{SL}(2,\mathbb{R})}{\mathrm{SO}(2)}\times \frac{\mathrm{SO}(2,n)}{%
\mathrm{SO}(2)\times \mathrm{SO}(n)}$, with $\mathcal{H}_{0}=\mathrm{SO}%
(2)\times \mathrm{SO}(n)$, $\mathbf{R}$ is reducible, being $\mathbf{R}=%
\mathbf{1}+\mathbf{n}$. In these cases we did not derive the explicit form
of the solutions to (\ref{SymplCond}), (\ref{M0QQ}) in terms of the
covariant building blocks defined above, and we leave this task for a future
investigation. Here, we limit ourselves to remark that, if we had the
explicit form for the solution $M_{MN}$ which reduces to $M_{+}$ once
truncated to the STU model, by the same token, since the STU truncation
comprises four charges in $\mathbf{n}+\bar{\mathbf{n}}$, the identification $%
\mathcal{M}^{H}=M$ would hold for the BPS solutions to these models, as well.

Notice that the above argument does not apply to the $\mathcal{N}>2$ models
in which the BPS solutions have non-trivial flat directions since, with
respect to the maximal compact subgroup $H_{0}$ of the stabilizer $G_{0}$ of
$\mathcal{Q}$ in $G$, the representation $\mathbf{R}$ in (\ref{RqR}) is
generally \textit{reducible}: $\mathbf{R}=\mathbf{R}_{1}+\mathbf{R}%
_{2}+\dots $. Moreover $\mathcal{M}^{H}$ depends on both $\mathcal{Q}$ and $%
\varphi _{flat}$, and thus it commutes with ${H}_{0}$ only at $\varphi
_{flat}=0$, being $H_{0}$ the stabilizer of this point. If however the
charges of the STU truncation belong to the $\mathbf{1}+\bar{\mathbf{1}}+%
\mathbf{R}_{1}+\bar{\mathbf{R}}_{1}$, we can at least state that $\mathcal{M}%
^{H}$, at $\varphi _{flat}=0$, and $M_{+}$ should coincide on the
corresponding subspace. Consider, for instance, the $\mathcal{N}=8$ theory.
In this case $G=\mathrm{E}_{7(7)}$, $G_{0}=\mathrm{E}_{6(2)}$, $H_{0}=%
\mathrm{SU}(2)\times \mathrm{SU}(6)$ and the representation $\mathbf{R}_{%
\mathcal{Q}}=\mathbf{56}$ branches as:
\begin{equation}
\mathbf{56}\,\,\,\overset{H_{0}}{\longrightarrow }\,\,\,\,\mathbf{1}+\mathbf{%
(1,15)}+\mathbf{(2,6)}+\bar{\mathbf{1}}+\overline{\mathbf{(1,15)}}+\overline{%
\mathbf{(2,6)}}\,,  \label{RqR2}
\end{equation}%
The charges of the STU truncation are in the $\mathbf{1}+\mathbf{(1,15)}+%
\bar{\mathbf{1}}+\overline{\mathbf{(1,15)}}$ and thus we expect $\mathcal{M}%
^{H}$, at $\varphi _{flat}=0$, and $M_{+}$ to coincide on these
representations, though not on the $\mathbf{(2,6)}+\overline{\mathbf{(2,6)}}$%
.\footnote{%
Although, at a generic point $\varphi _{flat}\neq 0$, $\mathcal{M}^{H}(%
\mathcal{Q},\varphi _{flat})$ does not commute with $H_{0}$, the matrix $%
\mathcal{S}^{H}(\mathcal{Q},\,\varphi _{flat})=\mathbb{C}\,\mathcal{M}^{H}(%
\mathcal{Q},\varphi _{flat})$ commutes with the group $H_{0}^{\prime }$
isomorphic to $H_{0}$ in $G_{0}$ and stabilizer of $\varphi _{flat}$ (in the
following we shall use the same symbol $H_{0}$ for the two isomorphic
subgroups of $G_{0}$). As a consequence of this, one can state on general
grounds that $\mathcal{S}^{H}(\mathcal{Q},\,\varphi _{flat})$ is
proportional to the identity on the irreducible representations of $%
H_{0}^{\prime }$ in the decomposition of $\mathbf{R}_{\mathcal{Q}}$.}

There is another notable property of both $M_{+}$ and $\mathcal{M}^{H}$
which is not shared by $M_{-}$: just as for $\mathcal{M}^{H}$, the adjoint
action of $M_{+}$ is an automorphism of $G$, namely
\begin{equation}
\left( M_{+}\right) ^{-1}\,\hat{R}_{\mathcal{Q}}[G]\,M_{+}\subset \hat{R}_{%
\mathcal{Q}}[G]\,\Leftrightarrow M_{+}\in {\mathrm{Aut}(G)},  \label{prr-1}
\end{equation}%
where $\hat{R}_{\mathcal{Q}}$ denotes the $2n\times 2n$ matrix
representation of $G$ in $\mathbf{R}_{\mathcal{Q}}$. The above property was
verified by computing the adjoint action of $M_{+}$ on the Lie algebra $%
\mathfrak{g}$ of $G$, in the representation $\mathbf{R}_{\mathcal{Q}}$, and
proving that it maps the algebra into itself.

%Using general properties of $M_+$, to be discussed below, we shall give, at
%the end of Sect. \ref{Summary}, an alternative argument in favor of the
%above identification.

Let us comment on the properties of the matrices $M_{\pm }$ under
Freudenthal duality $\mathfrak{F}$ (\ref{fd0}), and in particular under its
\textquotedblleft critical"/horizon version $\mathfrak{F}_{H}$ (\ref%
{FD-def-H}). By exploiting the properties of groups \textquotedblleft of
type $E_{7}$" \cite{brown}, one can show that
\begin{equation}
\mathfrak{F}_{H}\left( K_{MN}\right) =K_{MNPQ}\mathcal{\tilde{Q}}^{P}%
\mathcal{\tilde{Q}}^{Q}=K_{MN}-\frac{1}{6}\mathcal{\tilde{Q}}_{M}\mathcal{%
\tilde{Q}}_{N}+\frac{1}{6}\mathcal{Q}_{M}\mathcal{Q}_{N},  \label{f-I4>0}
\end{equation}%
which, in turn, implies
\begin{equation}
\mathfrak{F}_{H}(M_{\pm }(\mathcal{Q}))\equiv M_{\pm }(\mathfrak{F}_{H}(%
\mathcal{Q}))=M_{\pm }(\mathcal{Q})\,.  \label{FH1}
\end{equation}%
Thus, the identification (\ref{Res-5}) is consistent with the invariance of $%
\mathcal{M}_{MN}^{H}$ under $\mathfrak{F}_{H}$, as given Eq. (1.9) of \cite%
{FMY-FD-1}:
\begin{equation}
\mathfrak{F}_{H}\left( \mathcal{M}_{MN}^{H}\right) :=\mathcal{M}_{MN}^{H}(%
\mathcal{\tilde{Q}})=\mathcal{M}_{MN}^{H}(\mathcal{Q}).  \label{invv}
\end{equation}%
\begin{table}[h!]
\begin{center}
\begin{tabular}{|c||c|c|c|}
\hline
$%
\begin{array}{c}
\\
J_{3}%
\end{array}%
$ & $%
\begin{array}{c}
\\
G \\
~~%
\end{array}%
$ & $%
\begin{array}{c}
\\
\mathbf{R}_{\mathcal{Q}} \\
~~%
\end{array}%
$ & $%
\begin{array}{c}
\\
\left( d,n\right) \\
~~%
\end{array}%
$ \\ \hline\hline
$%
\begin{array}{c}
\\
J_{3}^{\mathbb{O}_{s}} \\
~%
\end{array}%
$ & $E_{7\left( 7\right) }~$ & $\mathbf{56}$ & $\left( 133,28\right) $ \\
\hline
$%
\begin{array}{c}
\\
J_{3}^{\mathbb{O}} \\
~%
\end{array}%
$ & $E_{7\left( -25\right) }~$ & $\mathbf{56}$ & $\left( 133,28\right) $ \\
\hline
$%
\begin{array}{c}
\\
J_{3}^{\mathbb{H}} \\
~%
\end{array}%
$ & $SO^{\ast }\left( 12\right) $ & $\mathbf{32}^{(^{\prime })}$ & $\left(
66,16\right) $ \\ \hline
$%
\begin{array}{c}
\\
J_{3}^{\mathbb{C}} \\
~%
\end{array}%
$ & $SU\left( 3,3\right) $ & $\mathbf{20}~$ & $\left( 35,10\right) $ \\
\hline
$%
\begin{array}{c}
\\
J_{3}^{\mathbb{R}} \\
~%
\end{array}%
$ & $Sp\left( 6,\mathbb{R}\right) $ & $\mathbf{14}^{\prime }$ & $\left(
21,7\right) ~$ \\ \hline
$%
\begin{array}{c}
\\
M_{1,2}\left( \mathbb{O}\right) \\
~%
\end{array}%
$ & $SU\left( 1,5\right) $ & $\mathbf{20}$ & $\left( 35,10\right) $ \\ \hline
$%
\begin{array}{c}
\\
\mathbb{R} \\
T^{3}~%
\end{array}%
$ & $SL\left( 2,\mathbb{R}\right) $ & $\mathbf{4}$ & $\left( 3,2\right) $ \\
\hline
$%
\begin{array}{c}
\\
\mathbb{R}\oplus \mathbb{R}\oplus \mathbb{R} \\
~STU%
\end{array}%
$ & $\left[ SL\left( 2,\mathbb{R}\right) \right] ^{3}$ & $\left( \mathbf{2},%
\mathbf{2},\mathbf{2}\right) $ & $\left( 9,4\right) $ \\ \hline
\end{tabular}%
\end{center}
\caption{Four-dimensional $U$-duality groups $G$, black hole charge
representation $\mathbf{R}_{\mathcal{Q}}$, and data $d:=$dim$\mathbf{Adj}$
and $n:=$dim$\mathbf{R}_{\mathcal{Q}}/2$. The corresponding scalar manifolds
are the \textit{symmetric} cosets $\frac{G}{H}$, where $H$ is the maximal
compact subgroup (with symmetric embedding) of $G$. $\mathbb{O}$, $\mathbb{H}
$, $\mathbb{C}$ and $\mathbb{R}$ respectively denote the four division
algebras of octonions, quaternions, complex and real numbers, and $\mathbb{O}%
_{s}$ is the split form of octonions. $M_{1,2}\left( \mathbb{O}\right) $ is
the Jordan triple system (not upliftable to $D=5$) generated by $2\times 1$
Hermitian matrices over $\mathbb{O}$ \protect\cite{GST}. Note that the $STU$
model \protect\cite{STU}, based on $\mathbb{R}\oplus \mathbb{R}\oplus
\mathbb{R}$, is reducible, but \textit{triality symmetric}. All cases
pertain to models with $8$ supersymmetries, with exception of $M_{1,2}\left(
\mathbb{O}\right) $ and $J_{3}^{\mathbb{O}_{s}}$, related to $20$ and $32$
supersymmetries, respectively. The $D=5$ uplift of the $T^{3}$ model based
on $\mathbb{R}$ is the \textit{pure} $\mathcal{N}=2$, $D=5$ supergravity. $%
J_{3}^{\mathbb{H}}$ is related to both $8$ and $24$ supersymmetries, because
the corresponding supergravity theories share the very same bosonic sector
\protect\cite{GST,twins}. All data $d$ and $n$ satisfy the relations (%
\protect\ref{ConsCond0})-(\protect\ref{ConsCond00}).}
\end{table}

Furthermore, the result (\ref{ABC-1}), as discussed in App. \ref{App-A}, is
constrained by the consistency condition%
\begin{equation}
d=\frac{3n(2n+1)\,}{n+8},  \label{ConsCond0}
\end{equation}%
relating the dimension $d$ of $G$ and the dimension $2n$ of the black hole
charge irrep. $\mathbf{R}_{\mathcal{Q}}$. As observed in \cite{Exc-Reds}, (%
\ref{ConsCond0}) actually characterizes \textit{at least} all the pairs $%
\left( G,\mathbf{R}_{\mathcal{Q}}\right) $ related to \textit{simple} rank-$%
3 $ Euclidean Jordan algebras \cite{GST} (such pairs are example of simple,
\textit{non-degenerate} groups \textquotedblleft of type $E_{7}$" \cite%
{FKM-Deg-E7}).

The cases related to $D=4$ Maxwell-Einstein gravity theories with local
supersymmetry are reported in Table 1; within this class, the so-called $STU$
model \cite{STU} is an exception: the corresponding rank-$3$ Jordan algebra
is \textit{semi-simple} ($\mathbb{R}\oplus \mathbb{R}\oplus \mathbb{R}$),
but however it still satisfies (\ref{ConsCond0}).

The condition (\ref{ConsCond0}) can be further elaborated, by observing
that, in all the cases under consideration, it holds that%
\begin{equation}
n=3q+4,
\end{equation}%
thus implying%
\begin{equation}
d=\frac{3(3q+4)\left( 2q+3\right) \,}{q+4}.  \label{ConsCond00}
\end{equation}%
For $J_{3}^{\mathbb{A}_{(s)}}$-related models (\textquotedblleft magical"
(super)gravities \cite{GST}), the parameter $q$ can be defined as
\begin{equation}
q:=\dim _{\mathbb{R}}\mathbb{A}_{(s)}=8,4,2,1~\text{for~}\mathbb{A}_{(s)}=%
\mathbb{O}_{(s)},\mathbb{H}_{(s)},\mathbb{C}_{(s)},\mathbb{R},
\end{equation}%
while $q=-2/3$ and $q=0$ for $T^{3}$ and $STU$ model, respectively (and $q=2$
for $\mathcal{N}=5$ theory).

\paragraph{Interpretation of $M_{-}$.}

Interestingly, also%
\begin{eqnarray}
M_{-,I_{4}>0\mid MN}(\mathcal{Q}) &=&\frac{4}{\left( I_{4}\right) ^{3/2}}%
\,K_{M}K_{N}-\frac{6}{\sqrt{I_{4}}}K_{MN}=\frac{1}{\sqrt{I_{4}}}\,\mathcal{%
\tilde{Q}}_{M}\mathcal{\tilde{Q}}_{N}-\frac{6}{\sqrt{I_{4}}}K_{MN}
\label{M_-} \\
&=&-\partial _{M}\partial _{N}\sqrt{I_{4}}  \label{M_-2}
\end{eqnarray}%
can be given a meaning within the stratification of $\mathbf{R}_{\mathcal{Q}%
} $ into $G$-orbits.

Indeed, $M_{-,I_{4}>0\mid MN}$ (\ref{M_-2}) can be regarded as the metric of
the non-compact pseudo-Riemannian rigid special K\"{a}hler manifold \cite%
{LG-2}%
\begin{equation}
\mathbf{M}_{I_{4}>0}:=\mathcal{O}_{I_{4}>0}\times \mathbb{R}^{+},  \label{s}
\end{equation}%
with real dimension $2n$; $\mathcal{O}_{I_{4}>0}$ denotes the corresponding
\textquotedblleft large" $G$-orbit defined by the $G$-invariant constraint $%
I_{4}>0$ on the charge representation $\mathbf{R}_{\mathcal{Q}}$ of $G$; the
$\mathbb{R}^{+}$ factor in (\ref{s}) simply corresponds to the non-vanishing
(strictly positive) values of $I_{4}$ itself. The signature along the $%
\mathbb{R}^{+}$-direction is negative, while the metric on $\mathcal{O}%
_{I_{4}>0}$ is that of the Cartan-Killing metric on the coset $G/G_0$, $G_0$
being the stabilizer of $\mathcal{Q}$, namely its positive and negative
eigenvalues correspond to the non-compact and compact generators in the
coset space, respectively.

In $\mathcal{N}=2$ (symmetric) theories, two $G$-orbits are defined by the
constraint $I_{4}>0$ : the ($\frac{1}{2}$-)BPS orbit, and the non-BPS $%
Z_{H}=0$ orbit \cite{BFGM-1}. Let us consider for instance the $\mathcal{N}%
=2 $ exceptional ``magical theory'' \cite{GST} ($G=E_{7(-25)}$, $\mathbf{R}_{%
\mathcal{Q}}=\mathbf{56}$), for which one can define the two
pseudo-Riemannian $56$-dimensional rigid special K\"{a}hler manifolds:%
\begin{eqnarray}
\mathbf{M}_{I_{4}>0,BPS} &:&=\mathcal{O}_{I_{4}>0,BPS}\times \mathbb{R}^{+}=%
\frac{E_{7(-25)}}{E_{6(-78)}}\times \mathbb{R}^{+}\text{, metric~}M_{-\mid
MN}\text{ with~}\left( n_{+},n_{-}\right) =\left( 54,2\right) ;  \notag \\
\mathbf{M}_{I_{4}>0,nBPS} &:&=\mathcal{O}_{I_{4}>0,nBPS}\times \mathbb{R}%
^{+}=\frac{E_{7(-25)}}{E_{6(-14)}}\times \mathbb{R}^{+}\text{, metric~}%
M_{-\mid MN}\text{ with~}\left( n_{+},n_{-}\right) =\left( 22,34\right) .
\notag \\
&&
\end{eqnarray}%
In general, the metric $M_{-\mid MN}$ of $\mathbf{M}_{I_{4}>0,BPS}$ always
has signature $\left( n_{+},n_{-}\right) =\left( 2n-2,2\right) $. This,
indeed, is nothing but the signature of the symplectic matrix $\mathcal{M}%
^{(F)}$, see (\ref{MFgen1}) or (\ref{MFgen2}) below, which will be proven in
Sect. \ref{Int-3} to coincide, for the BPS orbit, with $M_-$. In the example
of the STU truncation, for instance, one of the two positive eigenvalues of $%
M_{-}$ (\ref{M_-})-(\ref{M_-2}) is computed in App. \ref{App-B} for the
charge configuration $\left( q_{0},p^{1},p^{2},p^{3}\right) $, the other is
implied by $M_-$ being symplectic.

On the other hand, in the maximal $\mathcal{N}=8$ theory ($G=E_{7(7)}$, $%
\mathbf{R}_{\mathcal{Q}}=\mathbf{56}$) there is only one $G$-orbit defined
by the constraint $I_{4}>0$, namely the $\frac{1}{8}$-BPS \textquotedblleft
large" orbit, which thus allows to define the pseudo-Riemannian $56$%
-dimensional rigid special K\"{a}hler manifold \cite{LG-2}:%
\begin{equation}
\mathbf{M}_{I_{4}>0,\frac{1}{8}-BPS}:=\mathcal{O}_{I_{4}>0,\frac{1}{8}%
-BPS}\times \mathbb{R}^{+}=\frac{E_{7(7)}}{E_{6(2)}}\times \mathbb{R}^{+}%
\text{, metric~}M_{-\mid MN}\text{ with~}\left( n_{+},n_{-}\right) =\left(
30,26\right) .
\end{equation}%
\medskip

\subsection{\label{Int-2}Generalizing the Solutions $M_{\pm }$ to all $%
I_{4}\neq 0$ Orbits}

If we extend the expressions for $M_{\pm }$, given Section \ref{Simple}, to $%
I_{4}<0$:
\begin{align}
M_{+,I_{4}<0\,MN}& =\frac{1}{(-I_{4})^{\frac{3}{2}}}\left(
-8\,K_{M}K_{N}+6\,I_{4}\,K_{MN}-I_{4}\,\mathcal{Q}_{M}\mathcal{Q}_{N}\right)
\,,  \label{M+I4<0} \\
M_{-,I_{4}<0\,MN}& =\frac{1}{(-I_{4})^{\frac{3}{2}}}\left(
4\,K_{M}K_{N}-6\,I_{4}\,K_{MN}\right) \,.  \label{M_-2-I4<0}
\end{align}%
we find that, in contrast to the $I_{4}>0$ case, these matrices, though
still satisfying the condition (\ref{M0QQ}), are \emph{\ anti-symplectic},
namely satisfy the first of eq.s (\ref{meq2}) with $\epsilon=-1$. Under {the
}\textquotedblleft critical"/horizon version $\mathfrak{F}_{H}$ (\ref%
{FD-def-H}) of Freudenthal duality, $M_{\pm,I_{4}<0}$ transform as follows:
\begin{equation}
\mathfrak{F}_{H}(M_{\pm,I_{4}<0})=-M_{\pm,I_{4}<0}\,.  \label{invv-2}
\end{equation}%
This can be proved by using
\begin{equation}
\mathfrak{F}_{H}\left( K_{MN}\right) =K_{MNPQ}\mathcal{\tilde{Q}}^{P}%
\mathcal{\tilde{Q}}^{Q}=\epsilon \,K_{MN}-\frac{1}{6}\mathcal{\tilde{Q}}_{M}%
\mathcal{\tilde{Q}}_{N}+\frac{\epsilon }{6}\mathcal{Q}_{M}\mathcal{Q}_{N}\,,
\label{gen-prop}
\end{equation}
which generalizes (\ref{f-I4>0}) for any sign $\epsilon$ of $I_{4}$.
Correspondingly the properties (\ref{FH1}) and (\ref{invv-2}) can be
summarized as follows:
\begin{equation}
\mathfrak{F}_{H}(M_{\pm})=\epsilon\, M_\pm \,.  \label{freupm}
\end{equation}
As far as $M_{-}$ is concerned, for $I_4<0$, it coincides with the Hessian
of $-\sqrt{-I_4}$. As a consequence of this, in all regular orbits, we can
write, as a general property of $M_-$,
\begin{eqnarray}
M_{-,I_{4}>0\mid MN}(\mathcal{Q}) &=&-\partial _{M}\partial _{N}\sqrt{|I_{4}|%
}\,.  \label{M_-2-2}
\end{eqnarray}
Thus also for $I_4<0$, $M_-$ can be given the same interpretation as for the
$I_{4}>0$ case: $M_{-,I_{4}<0}$ can be regarded as the metric of the
non-compact pseudo-Riemannian rigid special K\"{a}hler manifold%
\begin{equation}
\mathbf{M}_{I_{4}<0}:=\mathcal{O}_{I_{4}<0}\times \mathbb{R}^{+},
\label{s-I4<0}
\end{equation}%
with real dimension $2n$; $\mathcal{O}_{I_{4}<0}$ denotes the unique
\textquotedblleft large" non-BPS $G$-orbit defined by the $G$-invariant
constraint $I_{4}<0$ on the charge representation $\mathbf{R}_{\mathcal{Q}}$
of $G$; the $\mathbb{R}^{+}$ factor in (\ref{s-I4<0}) simply corresponds to
the non-vanishing values of $|I_{4}|$ itself. For the $\mathcal{N}=2$
exceptional ``magical theory'' and $\mathcal{N}=8$ supergravity, the
manifold (\ref{s-I4<0}) is respectively given by%
\begin{align}
\mathcal{N}& =2:G=E_{7(-25)},\mathbf{R}_{\mathcal{Q}}=\mathbf{56}:\mathbf{M}%
_{I_{4}<0}:=\frac{E_{7(-25)}}{E_{6(-26)}}\times \mathbb{R}^{+}\text{,~metric~%
}M_{-\mid MN}\text{ with~}\left( n_{+},n_{-}\right) =\left( 28,28\right) ;
\notag \\
\mathcal{N}& =8:G=E_{7(7)},\mathbf{R}_{\mathcal{Q}}=\mathbf{56}:\mathbf{M}%
_{I_{4}<0}:=\frac{E_{7(7)}}{E_{6(6)}}\times \mathbb{R}^{+}\text{,~metric~}%
M_{-\mid MN}\text{ with~}\left( n_{+},n_{-}\right) =\left( 28,28\right) .
\notag
\end{align}%
Interestingly, the two manifolds share the same signature.

%
%An important difference between $M_{-,I_{4}>0}$ (\ref{M_-})-(\ref{M_-2}) and
%$M_{-.I_{4}<0}$ (\ref{M_-I4<0})-(\ref{M_-2-I4<0}) is that the former is
%symplectic, whereas the latter is \textit{anti-symplectic} (namely, it
%satisfies (\ref{anti-sympl})); indeed, by recalling that $\epsilon
%:=I_{4}/\left\vert I_{4}\right\vert $, it holds that%
%\begin{equation}
%M_{-}\mathbb{C}M_{-}=\epsilon \mathbb{C}.
%\end{equation}
As opposed to $M_{-}$, the adjoint action of $M_{+}$ defines, just as in the
$I_{4}>0$ case, an automorphism of $G$, namely satisfies eq. (\ref{prr-1}).
Since, however, for $I_{4}<0$ $M_{+}$ is antisymplectic, it can not be an
element of $G$, because the matrix realization $\hat{R}_{\mathcal{Q}}$ of
the elements of $G$ in the representation $\mathbf{R}_{Q}$ is symplectic. In
Appendix \ref{App-C} we argue that for \textquotedblleft type $\mathrm{E}%
_{7} $\textquotedblright\ supergravities the group $G$ has an outer
automorphism implemented by an antisymplectic matrix in the representation $%
\mathbf{R}_{\mathcal{Q}}$. Since, for $G$ simple, \emph{non-degenerate of
type E}$_{7}$, $\mathrm{Out}(G)=\mathrm{Aut}(G)/\mathrm{Inn}(G)$ has order
not greater than $2$ (see footnote 10 below), and its non-trivial element is
implemented by an antisymplectic matrix, a symplectic automorphism can only
be inner (see also footnote 12). We then conclude that, for $I_{4}>0$, $%
M_{+} $ defines an inner-automorphism, and is an element of $G$, while for $%
I_{4}<0 $ $M_{+}$ defines an outer-automorphism.

%
%
%for $I_{4}>0$ $M_{+}$ thus belongs to the \emph{inner}-automorphisms ${%
%\mathrm{Inn}(G)}\subset {\mathrm{Aut}(G)}$, whereas for $I_{4}<0$ the
%anti-symplecticity of $M_{+}$ implies that it belongs to the \emph{outer}%
%-automorphisms ${\mathrm{Aut}(G)}/{\mathrm{Inn}(G)}$
%\footnote{%
%Anticipating an argument to be given in Sect. \ref{Summary}, here we assume
%that $\mathrm{Out}(G)=\mathrm{Aut}(G)/\mathrm{Inn}(G)$ is contained in $%
%\mathbb{Z}_2$, as it seems to be common for groups of ``type $\mathrm{E}_{7}$%
%'' considered here. Since, as we shall prove in Appendix \ref{App-C}, the
%part of $\mathrm{Aut}(G)$ not connected to the identity is implemented by
%anti-symplectic matrices, symplectic automorphisms are inner.}, see Appendix %
%\ref{App-C}, because the matrix realization $\hat{R}_{\mathcal{Q}}$ of the
%elements of $G$ in the representation $\mathbf{R}_{Q}$ is symplectic:%
%\begin{equation}
%\hat{R}_{\mathcal{Q}}^{T}\mathbb{C}\hat{R}_{\mathcal{Q}}=\mathbb{C}.
%\end{equation}%
%The same generally does not hold for the adjoint action of $M_{-}$:%
%\begin{equation}
%\left( M_{-}\right) ^{-1}\,\hat{R}_{\mathcal{Q}}\,M_{-}\nsubseteq \hat{R}_{%
%\mathcal{Q}}\,\Leftrightarrow M_{-}\notin {\mathrm{Aut}(G)}.
%\end{equation}
We can define the matrix $S_+:=\mathbb{C}M_+$, which is still in $\mathrm{Aut%
}(G)$, since $M_+$ is. Moreover $S_+\mathcal{Q}=\mathbb{C}\mathcal{M}^H\,%
\mathcal{Q}=-\mathfrak{F}_H(\mathcal{Q})$. We can then use (\ref{freupm})
and write:
\begin{equation}
S_+^{-T}M_-(\mathcal{Q})S_+^{-1}=M_-(\mathfrak{F}_H(\mathcal{Q}%
))=\epsilon\,M_-(\mathcal{Q})\,,
\end{equation}
from which we can easily derive the following property:
\begin{equation}
M_+\mathbb{C}M_-\mathbb{C}M_+=-\epsilon\,M_-\,,
\end{equation}
or, equivalently:
\begin{equation}
M_-\,M_+^{-1}=M_+\,M_-^{-1}\,.  \label{mpmm}
\end{equation}
Finally it can be easily shown from their definition in both $I_4>0$ and $%
I_4<0$ cases, that
\begin{equation}
M_{\pm\,MN}\mathcal{Q}^{N}=\mathcal{M}_{MN}^{H}\mathcal{Q}^{N}=-\partial _{M}%
\sqrt{\left\vert I_{4}\right\vert }\,.  \label{rels-0}
\end{equation}

\subsection{\label{Int-3}Interpretation of $M_{\pm }$ in $\mathcal{N}=2$
Theories}

In the vector multiplet sector of an $\mathcal{N}=2$ supergravity, we can
define two symmetric, symplectic matrices: one is the matrix $\mathcal{M}$
constructed out of the real and imaginary parts of $\mathcal{N}_{\Lambda
\Sigma }$, as in (\ref{M-call}), the other is a matrix $\mathcal{M}^{(F)}$
defined by having the same matrix form as in (\ref{M-call}), but in terms of
the real and imaginary parts of the complex $n\times n$ matrix
\begin{equation}
\mathcal{F}_{\Lambda \Sigma }(X)=\frac{\partial ^{2}F}{\partial X^{\Lambda
}\partial X^{\Sigma }}\,,
\end{equation}%
$F(X)$ being the holomorphic prepotential, homogeneous function of degree $2$
of $X^{\Lambda }(z)$ (we use the notations of \cite{Andrianopoli:1996cm}).
We can write then:
\begin{align}
\mathcal{M}(z,\bar{z})& =\mathcal{M}[\mathrm{Re}\mathcal{N},\,\mathrm{Im}%
\mathcal{N}]\,, \\
\mathcal{M}^{(F)}(z,\bar{z})& =\mathcal{M}[\mathrm{Re}\mathcal{F},\,\mathrm{%
Im}\mathcal{F}]\,,
\end{align}%
where $\mathcal{M}[R,\,I]$ is the function of the matrices $R,\,I$ defined
in (\ref{M-call}). As anticipated in the introduction, can write the matrix $%
\mathcal{M}(z,\bar{z})$ in the manifestly symplectic-covariant form \cite%
{FK-N=8,Andrianopoli:2009je}
\begin{equation}
\mathcal{M}(z,\bar{z})=\mathbb{C}\left( V\bar{V}^{T}+\bar{V}V^{T}+U_{i}\,g^{i%
\bar{\jmath}}\bar{U}_{\bar{\jmath}}^{T}+\bar{U}_{\bar{\jmath}}g^{\bar{\jmath}%
i}U_{i}^{T}\right) \mathbb{C}\,.
\end{equation}%
Note that the right hand side is the sum of two symmetric matrices:
\begin{equation}
A_{1}=\mathbb{C}\left( V\bar{V}^{T}+\bar{V}V^{T}\right) \mathbb{C}%
\,\,;\,\,\,A_{2}=\mathbb{C}\left( U_{i}\,g^{i\bar{\jmath}}\bar{U}_{\bar{%
\jmath}}^{T}+\bar{U}_{\bar{\jmath}}g^{\bar{\jmath}i}U_{i}^{T}\right) \mathbb{%
C}\,,
\end{equation}%
which satisfy the condition $A_{1}\mathbb{C}A_{2}=0$, which follow from the
general properties: $V^{T}\mathbb{C}U_{i}=\bar{V}^{T}\mathbb{C}U_{i}=0$.
Therefore, if $\mathcal{M}=A_{1}+A_{2}$ is symmetric and symplectic, also $%
A_{1}-A_{2}$ is. The latter is just the matrix $\mathcal{M}^{(F)}$:
\begin{equation}
\mathcal{M}^{(F)}(z,\bar{z})=\mathbb{C}\left( V\bar{V}^{T}+\bar{V}%
V^{T}-U_{i}\,g^{i\bar{\jmath}}\bar{U}_{\bar{\jmath}}^{T}-\bar{U}_{\bar{\jmath%
}}g^{\bar{\jmath}i}U_{i}^{T}\right) \mathbb{C}\,,  \label{MFgen1}
\end{equation}%
The relation between the two matrices being then\footnote{%
This relation is also given in (1.13) of \cite{Ort-1}, in terms of the
so-called \textit{Hesse potential} (defined in (1.10) therein).}:
\begin{equation}
\mathcal{M}(z,\bar{z})=-\mathcal{M}^{(F)}(z,\bar{z})+2\,\mathbb{C}\left( V%
\bar{V}^{T}+\bar{V}V^{T}\right) \mathbb{C}\,,  \label{relationNF}
\end{equation}%
which is consistent with the relation between the lower diagonal blocks of
the two matrices given \textit{e.g.} in \cite{Ceresole:1995ca}:
\begin{equation}
\mathrm{Im}\mathcal{N}^{-1\,\Lambda \Sigma }=-\mathrm{Im}\mathcal{F}%
^{-1\,\Lambda \Sigma }-4\,L^{(\Lambda }\bar{L}^{\Sigma )}\,.
\end{equation}%
In $\mathcal{N}=2$ theories, we can express the matrix $\mathcal{M}^{(F)}$
in a form similar to Eq. (\ref{gen}) for $\mathcal{M}$, namely:
\begin{equation}
\mathcal{M}^{(F)}=-\mathbf{L}^{-T}\eta \mathbf{L}^{-1}\,,  \label{MFgen2}
\end{equation}%
where $\mathbf{L}$ is an $\mathrm{Sp}(2n,\mathbb{R})$-matrix of the form:
\begin{equation}
\mathbf{L}=\sqrt{2}\,\left( \mathrm{Re}(V),\,\mathrm{Re}({U}_{I}),-\mathrm{Im%
}(V),\,\mathrm{Im}(U_{I})\right) ;\;
\end{equation}%
moreover, $U_{I}=E_{I}{}^{i}U_{i}$, $E_{I}{}^{i}$ being the complex \textit{%
Vielbein} matrix of the special K\"{a}hler manifold, and $\eta $ is the
diagonal matrix:
\begin{equation}
\eta =\mathrm{diag}(1,\,-\mathbb{I}_{n-1},\,1,\,-\mathbb{I}_{n-1})\,,
\end{equation}%
where $\mathbb{I}_{n-1}$ denotes the $(n-1)\times (n-1)$ identity matrix.

Let us now evaluate relation (\ref{relationNF}) at the horizon of a regular
BPS black hole (thus, with $I_{4}>0$) and show that it yields the relation
between $M_{\pm }$, proving thus that, if $M_{+}$ coincides with the matrix $%
\mathcal{M}^{H}$, $M_{-}$ coincides with $\mathcal{M}^{(F)}$ at the horizon.
To this end, we use the relations \cite{FK-N=8}:
\begin{equation}
\left. 2i\,\bar{Z}\,V^{M}\right\vert _{\mathrm{horizon}}=\mathcal{Q}^{M}-i\,%
\mathbb{C}^{MN}\,\partial _{N}\sqrt{I_{4}}=\mathcal{Q}^{M}-\frac{2\,i}{\sqrt{%
I_{4}}}\,\mathbb{C}^{MN}\,K_{N}\,,
\end{equation}%
which hold at the horizon of the solution. Using the property that, at the
horizon, $|Z|_{\mathrm{horizon}}^{2}=\sqrt{I_{4}}$, we end up with
\begin{equation}
\left. 4\,V^{(M}\bar{V}^{N)}\right\vert _{\mathrm{horizon}}=\frac{1}{\sqrt{%
I_{4}}}\mathcal{Q}^{M}\mathcal{Q}^{N}+\frac{4}{\sqrt{I_{4}^{3}}}\,\mathbb{C}%
^{MP}\mathbb{C}^{NQ}\,K_{P}K_{Q}\,,
\end{equation}%
so that
\begin{equation}
\mathcal{M}^{H}=-\left. \mathcal{M}^{(F)}\right\vert _{\mathrm{horizon}}-%
\frac{1}{\sqrt{I_{4}}}\mathcal{Q}_{M}\mathcal{Q}_{N}-\frac{4}{\sqrt{I_{4}^{3}%
}}\,K_{M}K_{N}\,,  \label{relll-1}
\end{equation}%
which is the same relation holding between $M_{+}$ and $M_{-}$. Indeed, from
(\ref{genn-1}) and (\ref{genn-2}), it follows that%
\begin{equation}
M_{+\mid MN}=-M_{-\mid MN}-\epsilon \,\frac{1}{\sqrt{\left\vert
I_{4}\right\vert }}\mathcal{Q}_{M}\mathcal{Q}_{N}-\frac{1}{\sqrt{\left\vert
I_{4}\right\vert }}\mathcal{\tilde{Q}}_{M}\mathcal{\tilde{Q}}_{N}=-M_{-\mid
MN}-\epsilon \,\frac{1}{\sqrt{\left\vert I_{4}\right\vert }}\mathcal{Q}_{M}%
\mathcal{Q}_{N}-\frac{4}{\left\vert I_{4}\right\vert ^{3/2}}K_{M}K_{N},
\end{equation}%
which for $I_{4}>0$ reduces to the same relation (\ref{relll-1}).

\section{\label{Summary}General Discussion and Summary of Results}

We have constructed two symmetric real matrices $M_{\pm }(\mathcal{Q})$
satisfying the conditions (\ref{masterequation!}):
\begin{eqnarray}
M_{\pm }(\mathcal{Q})^{T}\mathbb{C}M_{\pm }(\mathcal{Q}) &=&\epsilon \,%
\mathbb{C}\,\,;\,\,  \label{p1} \\
\mathcal{Q}^{T}M_{\pm }(\mathcal{Q})\mathcal{Q} &=&-2\sqrt{|I_{4}|}\,,
\end{eqnarray}%
where $I_{4}=:\epsilon \,|I_{4}|$. These matrices also satisfy relations (%
\ref{rels-0}) :
\begin{equation}
M_{\pm\,MN}\mathcal{Q}^{N}=\mathcal{M}_{MN}^{H}\mathcal{Q}^{N}=-\partial _{M}%
\sqrt{\left\vert I_{4}\right\vert }\,.  \label{rels-2}
\end{equation}%
{The matrix}%
\begin{equation}
M_{-\mid MN}=\frac{4}{\left\vert I_{4}\right\vert ^{3/2}}K_{M}K_{N}-\epsilon
\frac{6}{\sqrt{\left\vert I_{4}\right\vert }}K_{MN}=\frac{1}{\sqrt{%
\left\vert I_{4}\right\vert }}\mathcal{\tilde{Q}}_{M}\mathcal{\tilde{Q}}%
_{N}-\epsilon \frac{6}{\sqrt{\left\vert I_{4}\right\vert }}K_{MN}=-\partial
_{M}\partial _{N}\sqrt{|I_{4}|},  \label{genn-2}
\end{equation}%
{which is never negative definite, enjoys an interpretation as symplectic
metric of the corresponding }$G$-orbit of $\mathcal{Q}$ (see above as well
as the final part of Sec. \ref{Simple}). Moreover it does \textit{not}
belong to {$\mathrm{Aut}(G)$}.

On the other hand, the matrix%
\begin{eqnarray}
M_{+\mid MN} &=&-\frac{8}{\left\vert I_{4}\right\vert ^{3/2}}%
K_{M}K_{N}+\epsilon \frac{6}{\sqrt{\left\vert I_{4}\right\vert }}%
K_{MN}-\epsilon \,\frac{1}{\sqrt{\left\vert I_{4}\right\vert }}\mathcal{Q}%
_{M}\mathcal{Q}_{N}  \label{pre-genn-1} \\
&=&-\frac{2}{\sqrt{\left\vert I_{4}\right\vert }}\mathcal{\tilde{Q}}_{M}%
\mathcal{\tilde{Q}}_{N}+\epsilon \frac{6}{\sqrt{\left\vert I_{4}\right\vert }%
}K_{MN}-\epsilon \,\frac{1}{\sqrt{\left\vert I_{4}\right\vert }}\mathcal{Q}%
_{M}\mathcal{Q}_{N}  \label{genn-1}
\end{eqnarray}
belongs to $\mathrm{Aut}(G)$ (in particular, see below, $M_{+,I_{4}>0}\in
\mathrm{Inn} (G)$ and $M_{+,I_{4}<0}\in {\mathrm{Aut}(G)}/\mathrm{Inn}(G)=:%
\mathrm{Out} (G)$; \textit{cfr. e.g.} App. \ref{App-C}).

Both matrices {under }$\mathfrak{F}_{H}$ (\ref{FD-def-H}){\ } transform as
in (\ref{freupm}).

For charges in a generic regular $G$-orbit (also in presence of flat
directions), one can construct the matrix:
\begin{equation}
\mathcal{A}(\mathcal{Q},\varphi _{flat}):=M_{+}(\mathcal{Q})^{-1}\,\mathcal{M%
}^{H}(\mathcal{Q},\varphi _{flat})\,,  \label{def-A}
\end{equation}%
so that%
\begin{equation}
\mathcal{M}^{H}(\mathcal{Q},\varphi _{flat})=M_{+}(\mathcal{Q})\mathcal{A}(%
\mathcal{Q},\varphi _{flat}).  \label{def-A-2}
\end{equation}

Let us illustrate some properties of $\mathcal{A}$; as it follows from from
Eq. (\ref{rels-2}), $\mathcal{A}(\mathcal{Q},\varphi _{flat})$ is in the
stabilizer of $\mathcal{Q}$ in $\mathrm{GL}(2n,\mathbb{R})$. Moreover, since
$M_{+}\in \mathrm{Aut}(G)$ and $\mathcal{M}^{H}\in G\subset \mathrm{Aut}(G)$%
, and both are invariant under ${H}_0$ (the stability group of $\varphi
_{flat}$), also $\mathcal{A}$ is, and thus we can write:
\begin{equation}
\mathcal{A}(\mathcal{Q},\varphi _{flat})\in \frac{\mathrm{Aut}(G)}{{H}_{0}}%
\cap \mathrm{Stab}_{\mathcal{Q}}[\mathrm{GL}(2n,\mathbb{R})]\,.  \label{Ain}
\end{equation}

An important property of $\mathcal{A}$ is the following:
\begin{equation}
\mathcal{A}^{T}\,M_{+}(\mathcal{Q})\mathcal{A}=M_{+}(\mathcal{A}^{-1}%
\mathcal{Q})=M_{+}(\mathcal{Q})\,,
\end{equation}%
which follows from (\ref{Ain}), but can be alternatively be proven using
Eq.s (\ref{def-A}), (\ref{ddd}), (\ref{p1}), and (\ref{freupm}):
\begin{align}
\mathcal{A}^{T}\,M_{+}(\mathcal{Q})\mathcal{A}& =\mathcal{M}^{H}\,M_{+}(%
\mathcal{Q})^{-1}\,\mathcal{M}^{H}=-\mathbb{C}\mathcal{S}^{H}M_{+}(\mathcal{Q%
})^{-1}\,(\mathcal{S}^{H})^{T}\mathbb{C}=-\mathbb{C}M_{+}(\mathcal{S}^{H}%
\mathcal{Q})^{-1}\mathbb{C}=  \notag \\
& =\epsilon \,M_{+}(\mathcal{S}^{H}\mathcal{Q})=\epsilon \,\mathfrak{F}%
_{H}(M_{+})=M_{+}(\mathcal{Q})\,.  \label{ampa}
\end{align}%
From this, it also follows that $\mathcal{A}$ is \textit{involutive}:
\begin{equation}
\mathcal{A}^{2}=\left( M_{+}\right) ^{-1}\mathcal{M}^{H}\,\left(
M_{+}\right) ^{-1}\mathcal{M}^{H}=\left( M_{+}\right) ^{-1}\,M_{+}=\mathbb{I}%
\,.
\end{equation}
Note that a property analogous to (\ref{ampa}) holds for $M_-$:
\begin{equation}
\mathcal{A}^{T}\,M_{-}(\mathcal{Q})\mathcal{A} =M_-\,,
\end{equation}
as it can be shown along the same lines as in (\ref{ampa}) and using
property (\ref{mpmm}).

If $I_{4}<0$, $M_{+}(\mathcal{Q})$ is \textit{anti-symplectic}, and thus (%
\ref{def-A}) yields that $\mathcal{A}$ is \textit{anti-symplectic} as well.
Therefore, as $M_{+}(\mathcal{Q})$, it defines an \textit{outer}%
-automorphism of $G$ (see Appendix \ref{App-C} for a discussion on
anti-symplectic outer-automorphisms of the U-duality algebra), and one can
write:
\begin{eqnarray}
M_{+}(\mathcal{Q}) &\in &\mathrm{Out}(G); \\
\mathcal{A}_{I_{4}<0}(\mathcal{Q},\varphi _{flat}) &\in &\mathrm{Out}(G)\cap
\mathrm{Stab}_{\mathcal{Q}}[\mathrm{GL}(2n,\mathbb{R})]\,.  \label{def-A-3}
\end{eqnarray}%
In the special case of the $T^{3}$-model the $I_{4}<0$ non-BPS solution has
no flat direction and thus $\mathcal{A}_{I_{4}<0}(\mathcal{Q})$ is a purely
charge dependent antisymplectic matrix in the stabilizer of $\mathcal{Q}$
\begin{equation}
\mathcal{M}_{\mathcal{I}_{4}<0}^{H}=M_{+}(\mathcal{Q})\,\mathcal{A}%
_{I_{4}<0}(\mathcal{Q}).
\end{equation}%
Note that, \textit{at least} in those cases\footnote{%
An interesting reference in which these properties of real forms of simple
Lie groups are listed is %
\url{http://en.wikipedia.org/wiki/List_of_simple_Lie_groups} (see also
references therein). We thank G. Dall'Agata for pointing it out to us.} in
which%
\begin{equation}
\mathrm{Out}(G)\subset \mathbb{Z}_{2},  \label{Aut/Inn}
\end{equation}%
which comprise all simple, \emph{non-degenerate type E}$_{7}$ groups $G$
(including thus $\mathrm{E}_{7(7)}$ itself) \cite{brown} in $D=4$
supergravity, all non-trivial outer-automorphisms are implemented by an
\textit{anti-symplectic} transformation.

If $I_{4}>0$, $M_{+}(\mathcal{Q})$ (\textit{cfr.} (\ref{p1})) is \textit{%
symplectic}, and thus (\ref{def-A}) yields that $\mathcal{A}$ is \textit{%
symplectic} as well. Therefore, as $M_{+}(\mathcal{Q})$, it defines an
\textit{inner}-automorphism of $G$, and one can write (with $\mathcal{Q}$
belonging to regular $G$-orbits with $I_{4}>0$; $H_0=\mathcal{H}%
_{0}=mcs\left( G\right) /U(1)$ in the BPS case, while, in the non-BPS case,
it is given for instance in \cite{BFGM-1}):
\begin{eqnarray}
M_{+}(\mathcal{Q}) &\in &\mathrm{Inn}(G)=G; \\
\mathcal{A}_{I_{4}>0}(\mathcal{Q},\varphi _{flat})\,&\in& \frac{G}{{H}_{0}}%
\cap \mathrm{Stab}_{\mathcal{Q}}[\mathrm{Sp}(2n,\mathbb{R})]\,.
\label{def-A-4}
\end{eqnarray}

In the absence of flat directions $\varphi _{flat}$ (such as for $\mathcal{N}%
=2$ regular BPS orbit), namely in those cases considered in Sec. \ref{Geom},
$G_{0}=H_{0}$, we have:
\begin{equation}
\frac{G}{{H}_{0}}\cap \mathrm{Stab}_{\mathcal{Q}}[\mathrm{Sp}(2n,\mathbb{R}%
)]=\frac{G_{0}}{H_{0}}=\{Id\}\,.
\end{equation}%
so that property (\ref{def-A-4}) implies
\begin{equation}
\mathcal{A}_{I_{4}>0}(\mathcal{Q},\varphi _{flat})=Id\,,
\end{equation}%
which is consistent with the identification $\mathcal{M}^{H}=M_{+}$ made in
Sect. \ref{Geom} (\textit{cfr.} (\ref{Res-5})).\bigskip
\paragraph{$M_+$ as a symmetry transformation.}
The property of $M_+$ of being an automorphism of $\mathfrak{g}$ implies its leaving the $K$-tensor invariant.
Indeed let $\{t'_\alpha\}$ denote the basis of $\mathfrak{g}$ resulting from an adjoint action of $M_+$ on $\{t_\alpha\}$. Being $M_+$ an automorphism we have:
\begin{equation}
t'_\alpha=M_+^{-1}\,t_\alpha\, M_+=M_{\alpha}{}^\beta\,t_\beta\,.
\end{equation}
This action clearly leaves the invariant tensor $\eta_{\alpha\beta}:={\rm Tr}(t_\alpha\,t_\beta)$ unaltered:
\begin{equation}
\eta_{\alpha\beta}:={\rm Tr}(t_\alpha\,t_\beta)={\rm Tr}(t'_\alpha\,t'_\beta)=M_{\alpha}{}^\gamma\,M_{\beta}{}^\delta\,\eta_{\gamma\delta}\,.
\end{equation}
As a consequence of this, using the general expression (\ref{K}) for $K_{MNPQ}$, we conclude that the $K$-tensor expressed in terms of $t_\alpha$ or $t'_\alpha$ coincide, i.e. that it is $M_+$-invariant.
If ${I}_4>0$, $M_+$ also leaves the symplectic form $\mathbb{C}$ invariant, and thus is an element of $\hat{R}_{\mathcal{Q}}[G]$, as previously emphasized. If, in the other hand, ${I}_4<0$, $M_+$, being anti-symplectic, does not leave $\mathbb{C}$ invariant, but can, nevertheless, be thought of as an element of  the space $\hat{R}_{\mathcal{Q}}[G]\cdot \mathcal{O}$, where $\mathcal{O}$ is the involutive anti-symplectic matrix defined in Appendix \ref{App-C}. In the former case ($I_4>0$) $M_+$  is a  charge-dependent symmetry of the theory while in the latter ($I_4<0$), the presence of $\mathcal{O}$ makes $M_+$ a symmetry only if combined with a parity or time-reversal transformation \cite{AT}. In both cases $M_+$, as opposed to $\mathcal{M}^H$ when $\varphi_{flat}\neq 0$, only depends on the charges. Although the actions of the two matrices $M_+$ and $\mathcal{M}^H$ coincide on $\mathcal{Q}$ (and define the Freudenthal dual), they differ on the other fields of the theory. \par
In any case $M_+$ can be characterized as a $\hat{R}_{\mathcal{Q}}[G]$-valued function for $I_4>0$, or $\hat{R}_{\mathcal{Q}}[G]\cdot \mathcal{O}$-valued function for $I_4<0$,  over the duality orbit of $\mathcal{Q}$.
Let us conclude with a few comments.

A special role in our discussion has been played by outer-automorphisms of
the U-duality algebra which are implemented by anti-symplectic
transformations. These should correspond, modulo $U$-dualities, to a
discrete symmetry of ungauged extended supergravities, see Appendix \ref%
{App-C}, which deserves a separate discussion \cite{AT}.

Finally it would be interesting to extend our analysis to ``small orbits''
of $\mathbf{R}_Q$, for which $I_4=0$. \appendix

\section*{Acknowledgments}

MT would like to thank Prof. A. J. Di Scala, of the Department of
Mathematical Sciences of Politecnico di Torino, for useful discussions. SF
would like to thank M. Floratos for many interesting discussions on black
hole properties related to this investigation.

The work of SF is supported by the ERC Advanced Grant no. 226455 \textit{%
SUPERFIELDS}.

He is on leave of absence from Department of Physics and Astronomy,
University of California, Los Angeles, USA.

The work of AM is supported in part by the FWO - Vlaanderen, Project No.
G.0651.11, and in part by the Interuniversity Attraction Poles Programme
initiated by the Belgian Science Policy (P7/37).

The work of EO is supported by Brazilian Ministry of Science and Tecnology -
National Council of Scientific Development (CNPq).

The work of MT was supported by the Italian MIUR-PRIN contract
2009KHZKRX-007 ``Symmetries of the Universe and of the Fundamental
Interactions''.

\section{\label{K-Tensor}The $K$-Tensor}

Let us consider a $D=4$ $U$-duality group $G$ of real dimension $d$, with
generators $t^{\alpha }$ in the adjoint representation ($\alpha =1,...,d$).
The Gaillard-Zumino \cite{GZ} symplectic maximal embedding%
\begin{equation}
G\subset \mathrm{Sp}\left( 2n,\mathbb{R}\right)\,\,\,;\,\,\,\,\, \mathbf{R}_{%
\mathcal{Q}}=\mathbf{2n}
\end{equation}%
is provided by ($M,N=1,...,2n$)
\begin{equation}
t_{MN}^{\alpha }:=t_{\phantom\alpha M}^{\alpha \phantom{M}P}\mathbb{C}%
_{PN}\,,  \label{DefGen}
\end{equation}%
defining the Cartan-Killing metric $k_{\alpha \beta }$ of $G$ as%
\begin{equation}
\left( t_{\alpha |M}^{\phantom{\alpha M}N}t_{\beta |N}^{\phantom{\beta|N}%
M}\right) \equiv k_{\alpha \beta },
\end{equation}%
so that $t_{\alpha |M}^{\phantom{\alpha M}N}t_{\phantom{\alpha}N}^{\alpha %
\phantom NM}=d$. The tensor $t^{\alpha }{}_{MN}$ is a singlet of $G$ and,
being the representation $\mathbf{R}_{\mathcal{Q}}$ symplectic, is symmetric
in its symplectic indices:
\begin{equation}
t^{\alpha }{}_{MN}=t^{\alpha }{}_{\left( MN\right) }.
\end{equation}
\textit{At least} for groups $G$ \textquotedblleft of type $E_{7}$" \cite%
{brown}, it is possible to construct the aforementioned rank-$4$ completely
symmetric invariant tensor, dubbed $K$-tensor \cite{Exc-Reds}:
\begin{equation}
\exists !K_{MNPQ}\equiv \mathbf{1}\in \left( \mathbf{R}_{\mathcal{Q}}\times
\mathbf{R}_{\mathcal{Q}}\times \mathbf{R}_{\mathcal{Q}}\times \mathbf{R}_{%
\mathcal{Q}}\right) _{s},
\end{equation}%
which can be generally defined as follows:%
\begin{eqnarray}
K_{MNPQ}\propto t_{(MN}^{\alpha }t_{\alpha |PQ)} &=&\frac{1}{3}\left(
t^{\alpha }{}_{MN}t_{\alpha |PQ}+t^{\alpha }{}_{MP}t_{\alpha |QN}+t^{\alpha
}{}_{MQ}t_{\alpha |PN}\right)  \notag \\
&=&\frac{1}{4!}\left( 8\,t^{\alpha }{}_{MN}t_{\alpha |PQ}+16\,t^{\alpha
}{}_{M(P}t_{\alpha |Q)N}\right) \,.  \label{K0}
\end{eqnarray}%
Needless to say, the prototype of groups \textquotedblleft of type $E_{7}$"
is $E_{7}$ itself (pertaining to $\mathcal{N}=8$ and $\mathcal{N}=2$
supergravity, in its real forms $E_{7(7)}$ and $E_{7\left( -25\right) }$,
respectively), with $\mathbf{R}_{\mathcal{Q}}=\mathbf{56}$. By following the
treatment of \cite{Exc-Reds}, one can prove that
%%%%%%%%%%%%%%%%%%%%%%%%%%%%%%%%%%%%%%%%%%%%%%%%%%%%%%%%%%%%%%%%%%%%%%%
%\be
%%%%%%%%%%%%%%%%%%%%%%%%%%%%%%%%%%%%%%%%%%%%%%%%%%%%%%%%%%%%%%%%%%%%%%%
%t^\alpha{}_{M(P}t_{\alpha|Q)N}\propto\mC_{M(P}\mC_{Q)N}\,,
%%%%%%%%%%%%%%%%%%%%%%%%%%%%%%%%%%%%%%%%%%%%%%%%%%%%%%%%%%%%%%%%%%%%%%%
%\ee
%%%%%%%%%%%%%%%%%%%%%%%%%%%%%%%%%%%%%%%%%%%%%%%%%%%%%%%%%%%%%%%%%%%%%%%
%yielding the following simplified expression for (\ref{K0})
%%%%%%%%%%%%%%%%%%%%%%%%%%%%%%%%%%%%%%%%%%%%%%%%%%%%%%%%%%%%%%%%%%%%%%
\begin{equation}
K_{MNPQ}=\xi \left[ t_{MN}^{\alpha }t_{\alpha |PQ}-\tau \,\mathbb{C}_{M(P}%
\mathbb{C}_{Q)N}\right] \,,  \label{covquan}
\end{equation}%
where the real constants $\xi $ and $\tau $ have been introduced; the latter
can be determined by imposing the skew-tracelessness condition $\mathbb{C}%
^{NP}K_{MNPQ}=0$, yielding \cite{Exc-Reds}%
\begin{equation}
\tau =\frac{d}{n(2n+1)}\,,  \label{tau}
\end{equation}%
whereas, by consistency with the definitions used in literature (\textit{cfr.%
} \cite{ADFMT-1}, taking into account the different normalization
conventions), $\xi $ is fixed as
\begin{equation}
\xi =-\frac{1}{6\tau }=-\frac{n(2n+1)}{6d}.
\end{equation}%
Thus, the following general expression for the $K$-tensor is obtained:
\begin{equation}
K_{MNPQ}=-\frac{n\left( 2n+1\right) }{6d}\left[ t_{MN}^{\alpha }t_{\alpha
|PQ}-\frac{d}{n(2n+1)}\mathbb{C}_{M(P}\mathbb{C}_{Q)N}\right] \,,  \label{K}
\end{equation}%
The formula (\ref{K}) will be relevant to many subsequent computations (most
of them reported in Appendix \ref{App-A}). By contracting the $K$-tensor
with four charge vectors $\mathcal{Q}$'s, one obtains the quartic $G$%
-invariant homogeneous polynomial $I_{4}$ \cite{ADF-fixed} (\ref{I4-def}) in
$\mathbf{R}_{\mathcal{Q}}$, which can therefore be rewritten as
\begin{equation}
I_{4}:=K_{MNPQ}\mathcal{Q}^{M}\mathcal{Q}^{N}\mathcal{Q}^{P}\mathcal{Q}^{Q}=-%
\frac{1}{6\tau }t_{MN}^{\alpha }t_{\alpha |PQ}\mathcal{Q}^{M}\mathcal{Q}^{N}%
\mathcal{Q}^{P}\mathcal{Q}^{Q}\,.  \label{I4}
\end{equation}

\section{\label{App-A}Computing the Coefficients $A$, $B$ and $C$}

We will here report the derivation of result (\ref{ABC-1}), which can
actually be obtained in (\textit{at least}) two equivalent ways.

\subsection{With the Invariant Tensor $S_{MQ}^{\protect\alpha \protect\beta %
} $...}

We start from the condition (\ref{M0QQ}), which can be easily recast as%
\begin{equation}
A+\epsilon \left( B-\frac{\left( 2\tau -1\right) }{6\tau }C\right) =-2\,.
\label{I4cond}
\end{equation}

On the other hand, the implementation of the symplectic condition (\ref%
{SymplCond}) requires some further manipulations. By exploiting (\ref{KKCC-1}%
), one can rewrite (\ref{SymplCond}) as
\begin{eqnarray}
&&\mathbb{C}_{MQ}=M_{MN}M_{PQ}\mathbb{C}^{NP}  \notag \\
&=&\frac{1}{\left\vert I_{4}\right\vert }\left[ B-C\frac{\left( 2\tau
-1\right) }{6\tau }\right] ^{2}K_{N[M}K_{Q]P}\mathbb{C}^{NP}  \notag \\
&&+\frac{1}{\left\vert I_{4}\right\vert }\xi \left\{
\begin{array}{l}
-\frac{1}{6}A\left[ B-C\,\frac{\left( 2\tau -1\right) }{6\tau }\right] \\
+\frac{1}{6}C\,\left( \tau -1\right) \left[ B-C\frac{\,\left( 2\tau
-1\right) }{6\tau }\right] \\
+\frac{1}{6}\epsilon AC\,\left( \tau -1\right)%
\end{array}%
\right\} K_{[M}\mathbb{C}_{Q]A}\mathcal{Q}^{A},  \label{SymplCondFinal}
\end{eqnarray}%
where the result (obtained by explicit computation)%
\begin{equation}
K_{N}K_{PQ}K_{M}\mathbb{C}^{NP}=\frac{I_{4}}{72\tau }K_{[M}\mathbb{C}_{Q]A}%
\mathcal{Q}^{A}\,=K_{N}K_{P[Q}K_{M]}\mathbb{C}^{NP}  \label{KKKC}
\end{equation}%
was used. The skew-trace of (\ref{SymplCondFinal}) yields
\begin{eqnarray}
2n &=&M_{MN}M_{PQ}\mathbb{C}^{NP}\mathbb{C}^{MQ}  \notag \\
&=&-\frac{1}{6}\,(2\tau -1)\left[ B-C\frac{\left( 2\tau -1\right) }{6\tau }%
\right] ^{2}  \notag \\
&&+\frac{1}{6}\left\{
\begin{array}{l}
-\frac{1}{6}A\left[ B-C\,\frac{\left( 2\tau -1\right) }{6\tau }\right] \\
+\frac{1}{6}C\,\left( \tau -1\right) \left[ B-C\,\frac{\left( 2\tau
-1\right) }{6\tau }\right] \\
+\epsilon \,AC\,\frac{\left( \tau -1\right) }{6\tau }%
\end{array}%
\right\} \,,  \label{SkewCondd}
\end{eqnarray}%
where the result%
\begin{equation}
K_{MN}K_{PQ}\mathbb{C}^{NP}\mathbb{C}^{MQ}=-\frac{(2\tau -1)}{6\tau }I_{4}\,
\label{SkewKKC}
\end{equation}%
has been taken into account.

Since the left hand side of Eq. (\ref{SymplCondFinal}) is skew-symmetric,
the only way to obtain from (\ref{SymplCondFinal}) a further constraint (not
proportional to the skew-trace condition (\ref{SkewCondd})) on the real
coefficients $A$, $B$ and $C$ is to single out the terms not proportional to
the symplectic metric $\mathbb{C}_{MQ}$ itself. Group theoretical arguments (%
\textit{cfr.} \textit{e.g.} App. C of \cite{Exc-Reds}) lead to the following
decomposition:%
\begin{equation}
K_{MN}K_{PQ}\mathbb{C}^{NP}=\frac{1}{18n}\,\frac{1}{6\tau }\,I_{4}\,\mathbb{C%
}_{MQ}-\frac{2}{9n}\,\frac{1}{6\tau }K_{[M}\mathbb{C}_{Q]A}\mathcal{Q}^{A}-%
\frac{1}{36\tau ^{2}}\,t_{\alpha |(A_{1}A_{2}}S_{M)(Q}^{\alpha \beta
}t_{\beta |A_{3}A_{4})}\mathcal{Q}^{A_{1}}\,\mathcal{Q}^{A_{2}}\mathcal{Q}%
^{A_{3}}\mathcal{Q}^{A_{4}},  \label{KKC0}
\end{equation}%
where $S_{MQ}^{\alpha \beta }$ is a $G$-invariant tensor, satisfying \cite%
{Exc-Reds}
\begin{equation}
S_{MQ}^{\alpha \beta }=S_{\left[ MQ\right] }^{\left( \alpha \beta \right)
};~~S_{MQ}^{\alpha \beta }\mathbb{C}^{MQ}=0,  \label{S-props-1}
\end{equation}%
and the result%
\begin{equation}
f_{\alpha \beta \gamma }t^{\alpha }{}_{(MA_{1}}t^{\beta
}{}_{A_{2})(A_{3}}t^{\gamma }{}_{A_{4}Q)}\mathcal{Q}^{A_{1}}\mathcal{Q}%
^{A_{2}}\mathcal{Q}^{A_{3}}\mathcal{Q}^{A_{4}}=0\,  \label{R-2}
\end{equation}%
has been used.

Using the irreducible decomposition
\begin{equation}
-\frac{1}{6\tau }t_{\alpha |(MN}t_{\beta |PQ}S_{M)Q}^{\alpha \beta }=%
\mathcal{A}\,K_{(MNPQ}\mathbb{C}_{R)S}\,  \label{dd}
\end{equation}%
(where $\mathcal{A}$ is a constant to be determined), one can prove that the
three terms in the right hand side of (\ref{KKC0}) are not independent. In
fact, the following relation holds:%
\begin{equation}
K_{[M}\mathbb{C}_{Q]A}\mathcal{Q}^{A}=\frac{1}{4}I_{4}\mathbb{C}_{MQ}+\frac{1%
}{4\mathcal{A}\tau }\,\,t_{\alpha |(A_{1}A_{2}}S_{M)(Q}^{\alpha \beta
}t_{\beta |A_{3}A_{4})}\mathcal{Q}^{A_{1}}\mathcal{Q}^{A_{2}}\mathcal{Q}%
^{A_{3}}\mathcal{Q}^{A_{4}}\,,  \label{Link}
\end{equation}%
thus implying (\ref{KKC0}) to reduce to%
\begin{equation}
K_{MN}K_{PQ}\mathbb{C}^{NP}=-\left( 1+\frac{1}{2n\mathcal{A}}\right) \frac{1%
}{36\tau ^{2}}\,t_{\alpha |(A_{1}A_{2}}S_{M)(Q}^{\alpha \beta }t_{\beta
|A_{3}A_{4})}\mathcal{Q}^{A_{1}}\mathcal{Q}^{A_{2}}\mathcal{Q}^{A_{3}}%
\mathcal{Q}^{A_{4}}\,.  \label{KKC1}
\end{equation}%
Therefore, the finite symplecticity condition (\ref{SymplCondFinal}) for $%
\mathcal{M}^{H}$ can be rewritten as follows:%
\begin{eqnarray}
\mathbb{C}_{MQ} &=&M_{MN}M_{PQ}\mathbb{C}^{NP}  \notag \\
&=&-\frac{1}{24\tau }\,\epsilon \,\left\{
\begin{array}{l}
\epsilon \,\tau A\left[ B-C\frac{\left( 2\tau -1\right) }{6\tau }\right] \\
+C\,\frac{\left( \tau -1\right) }{6}\left[ B-C\frac{\left( 2\tau -1\right) }{%
6\tau }\right] \\
-\epsilon \,\frac{1}{6}AC\,\left( \tau -1\right)%
\end{array}%
\right\} \mathbb{C}_{MQ}\,  \notag \\
&&-\frac{1}{16\mathcal{A}|I_{4}|\tau ^{2}}\,\,\left\{
\begin{array}{l}
\frac{2}{9}\left( \frac{1}{n}+2\mathcal{A}\right) \left[ B-C\frac{\left(
2\tau -1\right) }{6\tau }\right] ^{2} \\
+\epsilon \,\tau A\left[ B-C\frac{\left( 2\tau -1\right) }{6\tau }\right] \\
+C\,\frac{\left( \tau -1\right) }{6}\left[ B-C\frac{\left( 2\tau -1\right) }{%
6\tau }\right] \\
+\epsilon \,AC\,\frac{\left( \tau -1\right) }{6}%
\end{array}%
\right\} t_{\alpha |(A_{1}A_{2}}S_{M)(Q}^{\alpha \beta }t_{\beta
|A_{3}A_{4})}\mathcal{Q}^{A_{1}}\mathcal{Q}^{A_{2}}\mathcal{Q}^{A_{3}}%
\mathcal{Q}^{A_{4}}\,.  \notag \\
&&  \label{SymplCondd-1}
\end{eqnarray}%
It is clear that $t_{\alpha |(A_{1}A_{2}}S_{M)(Q}^{\alpha \beta }t_{\beta
|A_{3}A_{4})}$ contains $t_{\alpha |A_{1}A_{2}}S_{MQ}^{\alpha \beta
}t_{\beta |A_{3}A_{4}}$ which, due to (\ref{S-props-1}), is orthogonal to
(and thus independent of) the symplectic metric $\mathbb{C}_{MQ}$. Thus, the
related coefficient has to be set to zero. This argument leads to the
following independent conditions:%
\begin{eqnarray}
&&-\frac{\epsilon \,}{6\tau }\left\{ \epsilon \,\tau A\left[ B-C\frac{\left(
2\tau -1\right) }{6\tau }\right] +C\,\frac{\left( \tau -1\right) }{6}\left[
B-C\frac{\left( 2\tau -1\right) }{6\tau }\right] +\epsilon \,AC\,\frac{%
\left( \tau -1\right) }{6}\right\} =4\,;  \notag \\
&&  \label{SymplEq1} \\
&&-\frac{1}{9}\epsilon \,\left[ B-C\frac{\left( 2\tau -1\right) }{6\tau }%
\right] ^{2}=-4\,.  \label{SymplEq2}
\end{eqnarray}%
In these relations, the real constant $\mathcal{A}$ introduced in the
decomposition (\ref{dd}) has been set to
\begin{equation}
\mathcal{A}=\frac{1}{2}\left( 3\tau -\frac{1}{n}\right) \,.  \label{A-call}
\end{equation}%
The result (\ref{A-call}) can be achieved by noticing that, using (\ref{dd}%
), the following equation holds:%
\begin{equation}
K_{N}K_{[M}K_{Q]P}\mathbb{C}^{NP}=-\frac{1}{36\tau }\left( \frac{1}{n}+2%
\mathcal{A}\right) I_{4}K_{[M}\mathbb{C}_{Q]A}\mathcal{Q}^{A}\,.
\label{KKC3}
\end{equation}%
$K_{N}K_{[M}K_{Q]P}\mathbb{C}^{NP}$ can also be elaborated through explicit
computation, and the result is given by Eq. (\ref{KKKC}). By comparing the
skew-traces of (\ref{KKC3}) and (\ref{KKKC}), (\ref{A-call}) follows.

It should be stressed that Eqs. (\ref{SymplEq1}) and (\ref{SymplEq2}) are
consistent with the skew-tracelessness condition (\ref{SkewCondd}) \textit{%
iff} the relation (\ref{ConsCond0}) holds. This means that only two
conditions out of the three ones given by Eqs. (\ref{SkewCondd}), (\ref%
{SymplEq1}) and (\ref{SymplEq2}) are independent. The third independent
condition is given by (\ref{I4cond}).

Thus, the solutions of the resulting system of three independent conditions
on the coefficients $A,\,B$ and $C$ occurring in the \textit{Ansatz} (\ref%
{Ans-1}) read as follows:%
%%%%%%%%%%%%%%%%%%%%%%%%%%%%%%%%%%%%%%%%%%%%%%%%%%%%%%%%%%%%%%%%%%%%%%
\begin{equation}
A=-2\mp \,6\sqrt{\epsilon }\,,\quad B=\frac{6\left( 1-2\tau \mp \tau \sqrt{%
\,\epsilon \,}\right) }{(\tau -1)}\,,\quad C=-\frac{36\tau \left( 1\pm \sqrt{%
\,\epsilon \,}\right) }{\,(\tau -1)}\,.  \label{ABC}
\end{equation}%
Since $A$, $B$ and $C$ must be real, (\ref{ABC}) implies that the treatment
is consistent \textit{only} for $I_{4}>0\Leftrightarrow \epsilon =+1$. Then,
specifying $\epsilon =+1$, (\ref{ABC}) simplifies down to the final result (%
\ref{ABC-1}).

We also add that the results (\ref{Link}) and (\ref{KKC1}) yield%
\begin{equation}
K_{MN}K_{PQ}\mathbb{C}^{NP}=-\frac{1}{27\tau }\left( \frac{1}{n}+2\mathcal{A}%
\right) K_{[M}\mathbb{C}_{Q]A}\mathcal{Q}^{A}+\frac{1}{18}\left( \frac{1}{n}%
+2\mathcal{A}\right) \frac{1}{6\tau }I_{4}\mathbb{C}_{MQ}\,.  \label{KKC4}
\end{equation}%
Clearly, the skew-trace of the Eq. (\ref{KKC4}) must coincide with Eq. (\ref%
{SkewKKC}), thus implying the consistency condition (\ref{ConsCond0}).

\subsection{...and without $S_{MQ}^{\protect\alpha \protect\beta }$}

\label{withoutS}

By inserting (\ref{A-call}) into (\ref{KKC4}), one obtains

\begin{equation}
K_{MN}K_{PQ}\mathbb{C}^{NP}=-\frac{1}{9}K_{[M}\mathbb{C}_{Q]P}\mathcal{Q}%
^{P}+\frac{1}{36}I_{4}\mathbb{C}_{MQ}\,=-\frac{1}{9}K_{[M}\mathcal{Q}_{N]}+%
\frac{1}{36}I_{4}\mathbb{C}_{MQ},  \label{KKC5}
\end{equation}%
which, by further contracting with $\mathcal{Q}^{Q}$, yields%
\begin{equation}
K_{MN}K_{P}\mathbb{C}^{NP}=-K_{MP}\mathbb{C}^{NP}K_{N}=\frac{1}{12}I_{4}%
\mathcal{Q}_{M}.  \label{KKC6}
\end{equation}

Results (\ref{KKC5})-(\ref{KKC6}) actually hint for a simpler derivation of
result (\ref{ABC-1}), not involving of the use of the $G$-invariant tensor $%
S_{MQ}^{\alpha \beta }$ (\ref{S-props-1}) \cite{Exc-Reds} at all.

Indeed, starting from the \textit{Ansatz} (\textit{cfr.} (\ref{Ans-3}); $%
a,b,c\in \mathbb{R}$)%
\begin{equation}
M_{MN}(\mathcal{Q})=a\,K_{M}K_{N}+bK_{MN}+c\mathcal{Q}_{M}\mathcal{Q}_{N},
\end{equation}%
and observing that\footnote{%
Note that (\ref{R-1}) implies (\ref{R-2}).}%
\begin{equation}
-\frac{1}{2}f_{\alpha \beta \gamma }t_{MP}^{\alpha }t_{NQ}^{\beta
}t_{RS}^{\gamma }\mathcal{Q}^{P}\mathcal{Q}^{Q}\mathcal{Q}^{R}\mathcal{Q}%
^{S}=\tau ^{2}I_{4}\mathbb{C}_{MN}+2\tau ^{2}K_{[M}\mathcal{Q}_{N]},
\label{R-1}
\end{equation}%
after a little algebra Eqs. (\ref{KKC5})-(\ref{KKC6}) yield (\ref{ABC-1}):
\begin{equation}
\left\{
\begin{array}{l}
a=-\left( 2\pm 6\right) /\left\vert I_{4}\right\vert ^{3/2}; \\
\\
b=\pm 6/\left\vert I_{4}\right\vert ^{1/2}; \\
\\
c=-\left( 1\pm 1\right) /2\left\vert I_{4}\right\vert ^{1/2}.%
\end{array}%
\right.  \label{formaMdef}
\end{equation}

\section{\label{App-B}Signature of $M_{-}$}

In all $\mathcal{N}$-extended, $D=4$ supergravity theories based on
non-degenerate \cite{FKM-Deg-E7} $U$-duality groups $G$ \textquotedblleft of
type $E_{7}$" \cite{brown}, a generic charge vector $\mathcal{Q}$ in the $G$%
-repr. $\mathbf{R}_{\mathcal{Q}}$ can be $G$-transformed (through the action
of a suitable element $\hat{g}\in G$) into a charge vector $\mathcal{Q}_{0}$
whose non-vanishing entries are only the charges $q_{0}$ and $p^{i}$ ($%
i=1,2,3$), pertaining to the $STU$ model truncation in the special
coordinates' frame (recall the absence of flat directions):%
\begin{equation}
\mathcal{Q\rightarrow Q}_{0}=\hat{g}^{-1}\mathcal{Q}\Rightarrow M\left(
\mathcal{Q}\right) \rightarrow \hat{g}^{-T}M\left( \mathcal{Q}_{0}\right)
\hat{g}^{-1}.
\end{equation}%
In particular, the definiteness properties of $M\mathcal{\ }$are preserved
by the action of $G$.

In particular, one can consider $M_{-}\left( \mathcal{Q}\right) $, given by (%
\ref{Ans-3})-(\ref{Ans-4}) and (\ref{ABC-1}) in the branch \textquotedblleft
$-$". As discussed in Sec. \ref{Geom}, $M_{-}$ is nothing but the opposite
of the Hessian of $\sqrt{I_{4}}$ (with $I_{4}>0$):%
\begin{equation}
M_{-\mid MN}=-\partial _{M}\partial _{N}\sqrt{I_{4}}.
\end{equation}%
Thus, in order to study its definiteness, it suffices to analyze the signs
of its diagonal elements. In the $STU$ truncation under consideration, it
can be explicitly computed that the first diagonal element is strictly
positive ($I_{4}=q_{0}p^{1}p^{2}p^{3}>0$):%
\begin{equation}
M_{-\mid 00}=q_{0}^{2}\sqrt{q_{0}p^{1}p^{2}p^{3}}>0,
\end{equation}%
thus implying that $M_{-\mid MN}$ is not negative definite.

On the other hand, it can be calculated that $M_{+}\left( \mathcal{Q}\right)
$, given by (\ref{Ans-3})-(\ref{Ans-4}) and (\ref{ABC-1}) in the branch
\textquotedblleft $+$", is diagonal, with all strictly negative elements,
and thus trivially negative definite.

\section{\label{App-C} Outer (Anti-symplectic) Automorphisms of $\mathfrak{g}
$}

In symmetric extended $D=4$ supergravities, the U--duality algebra $%
\mathfrak{g}$ admits an automorphism implemented, in the representation $%
\mathbf{R}_{Q}$, by an anti-symplectic transformation. Consider the
symplectic frame in which the elements of a suitable basis of $\mathfrak{g}_4$ are represented, through  $\hat{R}_{\mathcal{Q}}$, either by matrices whose entries lie in the diagonal blocks or by matrices with entries only in the off-diagonal blocks. In this frame the conjugation by the
anti-symplectic matrix:
\begin{equation}
\mathcal{O}=\left(
\begin{matrix}
\mathbb{I}_{n} & \mathbf{0}_{n}\cr {\bf 0}_{n} & -\mathbb{I}_{n}%
\end{matrix}%
\right) \,,
\end{equation}%
defines an automorphism:
\begin{equation}
\mathcal{O}^{-1}\,\hat{R}_{\mathcal{Q}}[\mathfrak{g}]\,\mathcal{O}=\hat{R}_{%
\mathcal{Q}}[\mathfrak{g}]\,,
\end{equation}
where $\hat{R}_{\mathcal{Q}}[\mathfrak{g}]$ denotes the algebra of all
symplectic matrices representing $\mathfrak{g}$. For instance, in the
maximal theory, such transformation switches the sign of the generators in
the $\mathbf{35}_{c}$ (parametrized by the pseudo-scalars) and $\mathbf{35}%
_{s}$ (compact generators in $\mathfrak{su}(8)\ominus \mathfrak{so}(8)$),
leaving the other generators unaltered \cite{Trig-2}.

Since all $G$ transformations in $\mathbf{R}_{\mathcal{Q}}$ are implemented
by symplectic matrices, $\mathcal{O}$ is not in $G$ and defines a
non-trivial outer automorphism\footnote{%
Strictly speaking, to show that $\mathcal{O}$ is an outer-automorphism, one
should prove that no other element of $G$ can induce the same transformation
on $\mathfrak{g}$. This is immediate if $\mathbf{R}_{\mathcal{Q}}$ is
irreducible since any other real matrix inducing the same transformation,
must be proportional to $\mathcal{O}$, and thus non-symplectic. Inspection
of supergravities in which $\mathbf{R}_{\mathcal{Q}}$ is reducible, however,
leads to the same conclusion: No element of $G$ can induce the same
automorphism as $\mathcal{O}$.} \footnote{%
The simplest example of a real Lie group admitting a symplectic
representation in which an outer automorphism is implemented by an
anti-symplectic transformation, is $\mathrm{SL}(2,\mathbb{R})$ : its
fundamental representation $\mathbf{2}$ is symplectic and the
anti-symplectic matrix $\sigma _{3}=\mathrm{diag}(+1,-1)$ (which corresponds
to the limit $n=1$ in (\ref{D.5})) implements an outer-automorphism. The
same holds for the spin $3/2$ representation $\mathbf{4}$ (with the
anti-symplectic outer-automorphism given by (\ref{D.5}) with $n=2$), which
also characterizes $\mathrm{SL}(2,\mathbb{R})$ as the simplest example of
\emph{non-degenerate} group \emph{of type E}$_{7}$.} of $\mathfrak{g}$:
\begin{equation}
\mathcal{O}\in \frac{\mathrm{Aut}(G)}{\mathrm{Inn}(G)}=\mathrm{Out}(G)\,.
\end{equation}

We can give an alternative representation of $\mathcal{O}$, for those
supergravities admitting a $D=5$ uplift, in the symplectic frame originating
from the $D=5\rightarrow D=4$ reduction. These class of models comprises all
\textquotedblleft type $\mathrm{E}_{7}$\textquotedblright\ supergravities,
excluded the \textquotedblleft degenerate\textquotedblright\ ones, see
footnote 7. In this frame the generators $t_{\alpha }$ of $\mathfrak{g}$
have a characteristic matrix form given in \cite{Gnecchi-3}, defined by
branching the $D=4$ duality algebra with respect to $\mathrm{O}(1,1)\times
G_{5}$, $G_{5}$ being the global symmetry group of the $D=5$ parent theory.
The algebra $\mathfrak{g}$ decomposes accordingly:
\begin{equation}
\mathfrak{g}=[\mathfrak{o}(1,1)\oplus \mathfrak{g}_{5}]_{0}\oplus \lbrack
\mathbf{R}_{-2}+\overline{\mathbf{R}}_{+2}]\,,  \label{decc}
\end{equation}%
where the subscripts refer to $\mathrm{O}(1,1)$-gradings, $\mathbf{R},\,%
\overline{\mathbf{R}}$ are $(n-1)$-dimensional (Abelian) spaces of nilpotent
generators transforming in the representations $\mathbf{R}$ and $\overline{%
\mathbf{R}}$ under the adjoint action of $G_{5}$, respectively. Generators
of $\mathfrak{g}$ in each of the subspaces on the right-hand-side of (\ref%
{decc}), have the following matrix form in $\mathbf{R}_{\mathcal{Q}}$:
\begin{align}
D& \in \mathfrak{o}(1,1)\,\,;\,\,\,D=\mathrm{diag}(-3,-\mathbb{I}_{n-1},3,%
\mathbb{I}_{n-1})\,,  \notag \\
\mathbf{E}(\lambda )& \in \mathfrak{g}_{5}\,\,;\,\,\,\mathbf{E}(\lambda )=%
\mathrm{diag}(1,\mathcal{E}(\lambda ),1,-\mathcal{E}(\lambda )^{T})\,,
\notag \\
T(a^{I})& \in \overline{\mathbf{R}}_{+2}\,\,;\,\,\,T(a^{I})=a^{I}\,T_{I}=%
\left(
\begin{matrix}
0 & 0 & 0 & 0\cr a^{J} & 0 & 0 & 0\cr0 & 0 & 0 & -a^{I}\cr0 & d_{IJ} & 0 & 0%
\end{matrix}%
\right) \,,  \notag \\
\bar{T}(b_{I})& \in \mathbf{R}_{-2}\,\,;\,\,\,\bar{T}(b_{I})=b_{I}%
\,(T_{I})^{T}\,,  \notag
\end{align}%
where $\mathcal{E}(\lambda )$ are $(n-1)\times (n-1)$ matrices representing
the generic element $\mathbf{E}(\lambda )$ of $\mathfrak{g}_{5}$. In this
basis the matrix there is the following anti-symplectic automorphism $%
\mathcal{O}$:
\begin{equation}
\mathcal{O}=\left(
\begin{matrix}
1 & 0 & 0 & 0\cr0 & -\mathbb{I}_{n-1} & 0 & 0\cr0 & 0 & -1 & 0\cr0 & 0 & 0 &
\mathbb{I}_{n-1}%
\end{matrix}%
\right) \,,  \label{D.5}
\end{equation}%
whose action on the $\mathfrak{g}$-generators is:
\begin{equation}
\mathcal{O}^{-1}D\mathcal{O}=D\,\,;\,\,\,\mathcal{O}^{-1}\mathbf{E}(\lambda )%
\mathcal{O}=\mathbf{E}(\lambda )\,\,;\,\,\,\mathcal{O}^{-1}T(a^{I})\mathcal{O%
}=-T(a^{I})\,\,;\,\,\,\mathcal{O}^{-1}\bar{T}(a^{I})\mathcal{O}=-\bar{T}%
(a^{I})\,.
\end{equation}%
The anti-symplectic automorphism $\mathcal{O}$ is relevant for defining the $%
CP$-transfromation in supergravity \cite{AT}.

\end{document}